\documentclass[a4paper,11pt]{article}
\usepackage{jcappub} 
\usepackage{lineno}
\usepackage{float}
\usepackage{amsmath,amssymb,gensymb,graphicx,geometry,acronym}
\usepackage[symbol]{footmisc}
\usepackage{subcaption,array,multirow,cleveref,tabularray,longtable}

\makeatletter
\AtBeginDocument
 {
   \def\ltx@label#1{\cref@label{#1}}
   \def\label@in@display@noarg#1{\cref@old@label@in@display{#1}}
\def\label@in@mmeasure@noarg#1{%
    \begingroup%
      \measuring@false%
      \cref@old@label@in@display{#1}
    \endgroup}%  
 } %
\makeatother

\acrodef{GW}{gravitational wave}
\acrodef{SGWB}{stochastic gravitational wave background}
\acrodef{PTA}{pulsar timing array}
\acrodef{CMB}{cosmic microwave background}
\acrodef{LVK}{LIGO/Virgo/KAGRA}
\acrodef{INDIGO}{Indian Initiative in Gravitational-Wave Observations}
\acrodef{ET}{Einstein Telescope}
\acrodef{CE}{Cosmic Explorer}
\acrodef{LISA}{Laser Interferometer Space Antenna}
\acrodef{DECIGO}{Deci-hertz Interferometer Gravitational wave Observatory}
\acrodef{BBO}{Big Bang Observer}
\acrodef{IPTA}{International Pulsar Timing Array}
\acrodef{NANOGrav}{North American Nanohertz Observatory for Gravitational Waves}
\acrodef{EPTA}{European Pulsar Timing Array}
\acrodef{PPTA}{Parkes Pulsar Timing Array}
\acrodef{InPTA}{Indian Pulsar Timing Array}
\acrodef{FAST}{Five-hundred-meter Aperture Spherical Telescope}
\acrodef{SKA}{Square Kilometre Array}
\acrodef{SMBHBs}{supermassive black hole binaries}
\acrodef{WD}{white dwarf}
\acrodef{SOBH}{stellar origin black hole}
\acrodef{NS}{neutron star}
\acrodef{EMRI}{extreme mass ratio inspiral}
\acrodef{PBH}{primordial black hole}
\acrodef{BICEP}{Background Imaging of Cosmic Extragalactic Polarisation}
\acrodef{FOPT}{first order phase transition}
\acrodef{CPTA}{Chinese Pulsar Timing Array}
\acrodef{SFSR}{single field slow roll}
\acrodef{PSD}{power spectral density}
\acrodef{SNR}{signal-to-noise ratio}
\acrodef{nwt}{`noise-with-time'}
\acrodef{PLISC}{power law integrated sensitivity curve}

\arxivnumber{}
\title{\boldmath How to Constrain the Stochastic Gravitational Wave Background with Multi-Frequency Detections}

\author[1]{E. Gleave\note{Corresponding author.}}
\author{and A. H. Jaffe}
\affiliation{Imperial College London,\\Blackett Laboratory, Prince Consort Road, UK}

\emailAdd{e.gleave20@imperial.ac.uk}

\abstract{
Gravitational wave (GW) observations probe both a diffuse, stochastic gravitational wave background (SGWB) as well as individual cataclysmic events such as the merger of two compact objects. The detection and description of the gravitational-wave background requires somewhat different techniques than required for individual events. In this paper, we probe the sensitivity of present and future GW telescopes to different background sources, including both those expected from unresolved compact binaries in both their quasi-Newtonian quiescent and their eventual mergers, as well as more speculative cosmological sources such as inflation, cosmic strings, and phase transitions, over regions in which those sources can be described by a single power law. We develop a Fisher matrix formalism to forecast coming sensitivities of single and multiple experiments, and novel visualizations taking into account the increase in sensitivity to a background over time. 
}

\begin{document}
\maketitle
\flushbottom

\section{Introduction}

In July 2023, the NANOGrav collaboration released their 15-year data, with the first confirmed observation of the \ac{SGWB} \cite{NG15_pulsars,NG15_gwb}. This has since been corroborated by other \acp{PTA} \cite{EPTA_gwb,PPTA_gwb,CPTA_gwb}, and we expect  additional detections over a range of frequencies from other experiments in the upcoming decades.

The \ac{SGWB} observed in the \ac{PTA} frequency range ($\sim10^{-9}$ Hz) could be the result of any number of cosmological phenomena, an astrophysical foreground of inspiralling \ac{SMBHBs} \cite{NG15_smbhbconstraints}, or indeed some combination. Cosmological sources include inflation (both immediate tensor perturbations \cite{Guzzetti}, and \acp{GW} associated with \acp{PBH} \cite{Domenech,yuan2021}, \acp{FOPT} \cite{hindmarsh_2014,Caprini_2016,Caprini_2020}, cosmic strings \cite{Kibble_1976}, and other more exotic phenomena including a variety of dark matter candidates \cite{Li_2017,Lambiase_2023,chowdhury2024}. At higher frequencies, we could also see contributions from any of these cosmological phenomena, and foregrounds from binary populations of different masses. Some of the cosmological phenomena are part of the $\Lambda$CDM model of cosmology, while others extend it, and some entail new physics. As such, observations of the \ac{SGWB} over a wide range of frequencies will have implications for our models of cosmology.

There are multiple avenues for analyzing contributions to  the \ac{SGWB}---some individual phenomena have physical constraints which narrow down their possible contributions to the \ac{SGWB}, and whatever \ac{SGWB} signal was propagating at recombination will leave imprints on the $B$-mode polarization of the \ac{CMB}.
But \ac{GW} experiments remain our most comprehensive method of observing the whole \ac{SGWB} spectrum across a wide frequency range.

Ultimately we hope to move from these initial narrow observations of the \ac{SGWB} to a wider spectrum. Current and planned \ac{GW} experiments span more than 12 decades of frequency. Figure~\ref{Fi:GW_sens} shows the current and predicted sensitivities of the \ac{GW} experiments considered in this paper, and shows the large gap in observable frequency between \acp{PTA} and space-based detectors; a range of possible intermediate-frequency experiments have been suggested (see, e.g., references \cite{kachelriess2024,Blas_2022,martens2023} for discussions) but none are yet expected to be implemented. Many of these experiments will not come online for over a decade, and it is useful to know in advance which are likely to be useful for observing which phenomena \textit{before} we have access to their results. This allows us to establish, for example, whether we should wait for \ac{GW} results or prioritise looking for additional physical constraints in the interim.

Previous papers (e.g., \cite{Gowling_2021,tan2024}) have considered the predicted sensitivities of a single experiment to one, or sometimes multiple, phenomena; other papers have considered the joint sensitivity of two similar experiments to specified astrophysical foregrounds (e.g., \cite{Torres_Orjuela_2024,torresorjuela2024}). In this paper we consider which combinations of experiments will provide better sensitivity when evaluated collectively---many of the possible \ac{SGWB} spectra predicted lie across multiple experiments' frequency ranges. We generate results which can be easily applied to a range of phenomena, by focusing on the sensitivities of individual and combinations of experiments to generic spectra. These generic spectra can be fit to analytical or simulated spectra from different phenomena to get sensitivities to the physics involved.

All of the cosmological phenomena mentioned above have spectra which can be reasonably fit by single or broken power laws---See figure~\ref{Fig:example_spectra} for a range of example spectra. This paper looks only at single (`straight line') power laws. While there are examples of single power laws, broken power laws are often a more accurate model for the \ac{GW} spectra considered. However, single power laws by definition make up the straight sections of broken power laws, and it is useful to consider these separately. It is also worth noting that often broken power laws are often indistinguishable from single power laws across the frequency band of a single experiment. We extend the formalism developed in this paper to broken power laws, and hence a wider variety of possible phenomena in an upcoming paper \cite{paper2}.

However the primary reason for considering only single power laws in this paper is that it allows for a simple diagonalisation of the covariance matrix. That is, we establish a `pivot point' in frequency for each scenario where the errors on the slope and amplitude are decorrelated and can be analysed individually.

By using Fisher matrices, we analyse the sensitivities of various \ac{GW} experiments, focusing on the increase in sensitivity of multiple experiments when considered in parallel. This paper builds on the work of Gowling and Hindmarsh \cite{Gowling_2021} 
(which used Fisher matrices to determine the sensitivity of LISA alone to first-order phase transitions, modelled as a specific subset of double broken power laws) to consider how well different individual \ac{GW} experiments, and various combinations of experiments, will inform us about the contributing phenomena of the \ac{SGWB}. This analysis is carried out with foreground contamination, modelled here with an adapted and universal model across each population. We also explicitly investigate how successfully gaps in observing frequency can be addressed, considering combinations of experiments which lie on either side of the gap.

The Fisher matrix formalism also enables us to include a prior distribution on the fiducial models tested, given in this case by the posterior probability from the most recent NANOGrav analysis \cite{NG15_gwb,NG15_smbhbconstraints,NG15_newphysics}. We carry this out for both fiducial models which match the NANOGrav results, and for differing models. Where results are plotted, they are shown as novel `bowtie' regions, with a reasonable approximation to the true combination of errors on slope and amplitude.

Section~\ref{Se:Background} of the paper covers the relevant background, with information about the \ac{GW} experiments considered and examples of the expected \ac{GW} spectra that different cosmological phenomena and astrophysical foregrounds might contribute to the \ac{SGWB}. Current observations of and constraints on the \ac{SGWB} are discussed in Section~\ref{Se:Constraints}. Section~\ref{Se:Methodology} covers the analysis carried out, including the Fisher matrix set-up, and power law spectral model. Full noise and foreground models for the experiments considered are given in Appendices~\ref{A:noise} and~\ref{A:foreground}, respectively. Results are shown in Section~\ref{Se:Results}, and discussed in Section~\ref{Se:Discussion}.

\section{Background}\label{Se:Background}
\subsection{The SGWB}

The power spectrum of a \ac{GW} background can be described by its contribution to the energy density of the Universe per logarithmic frequency interval:
\begin{equation}
\label{eq:omega_h}
    \Omega_\textrm{GW}(f) = \frac{2\pi^2f^2}{3H_0^2}h_c^2(f)\;,
\end{equation}
where $h_c(f)$ is the characteristic strain at frequency $f$, and $H_0$ is the Hubble constant. Other equivalent descriptions of the energy density are given in section~\ref{SSec:modelling}. We can likewise define an equivalent effective energy density for instrumental noise, $\Omega_\textrm{ins}$ (discussed in more detail in section~\ref{SSSec:Ins_Noise}).

For definiteness, we will define the power law spectral index $\nu$ in terms of $\Omega_\textrm{GW}(f)\propto f^{\nu}$ (see eq.~\ref{eq:powerlaw}).

\subsection{Gravitational Wave Experiments}

\begin{figure}
\centering
\includegraphics{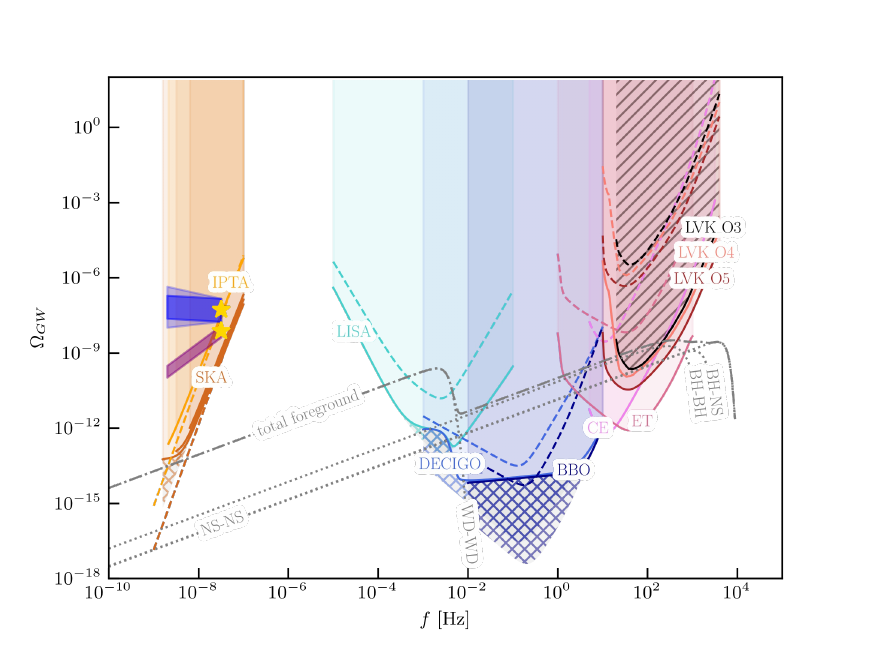}
\caption{Current results and eventual sensitivities for the experiments considered in this paper. Dashed lines show the instantaneous noise curves $\Omega_\textrm{nominal}$, detailed in appendix~\ref{A:noise}. Solid lines show the eventual noise curves $\Omega_\textrm{final}$ which take into account the improvement of sensitivity with observing time and the presence of any foregrounds. The cross hatched regions show the eventual sensitivities in the absence of any foregrounds---see section~\ref{SSSec:nwt} for discussion of $\Omega_\textrm{noise with time}$ and its difference to standard power law integrated sensitivity curves (PLISCs). Note that $\Omega_\textrm{final}$ for SKA is shown after 5/10/20 years, and see section~\ref{SSSec:nwt} for the issues with $\Omega_\textrm{nominal}$ for PTAs. The diagonally hatched region shows the upper limit given by LVK-O3's lack of SGWB detection. The most recent NANOGrav results \cite{NG15_gwb} are shown by the blue and purple shaded regions, and are discussed in more detail in figure~\ref{subfig:NG1}. Dotted lines show the predicted instantaneous astrophysical foregrounds from different binary populations, using equation~\ref{eq:FG_tanh} fitted to models for white dwarf \cite{Carter_2012}, and black hole, neutron star and black hole-neutron star binaries \cite{LVK_2021,LVK_2021b}.}
\label{Fi:GW_sens}
\end{figure}

Different sizes of \ac{GW} detector are sensitive to different frequencies of gravitational waves---the larger the arm length of the detector, the lower the frequencies it can reach, as can be seen in figure~\ref{Fi:GW_sens}. More details and references for each experiment considered in this work are given in table~\ref{Ta:GWexps}.

Ground-based detectors are restricted to kilometres in size. The currently active ground-based detectors are all involved in the \ac{LVK} collaboration, currently in its fourth operation phase, LVK-O4. Future ground-based detectors include upgrades to \ac{LVK}, as well as new experiments including \ac{INDIGO}, the \ac{ET}, and \ac{CE}. \ac{ET} and \ac{CE} are intended to work in conjunction with one another, primarily to increase the signal-to-noise and improve localisation of individual sources on the sky \cite{evans2021}.

Much larger space-based interferometers will be constructed, with arm lengths up to scales of millions of kilometres. \ac{LISA} is one such project, while \ac{DECIGO} and \ac{BBO} have also been proposed as additional space-based detectors. 

The lowest frequency detectors we have are \acp{PTA}. These exploit the properties of pulsars to set up effective interferometers with arms which stretch as far away as the pulsars themselves. Millisecond pulsars are sensitive to gravitational waves between $10^{-9}$ and $10^{-6}$ Hz. Active \acp{PTA} include those in the \ac{IPTA} collaboration (NANOGrav, the \ac{EPTA}, the \ac{PPTA}, the \ac{InPTA}) and the \ac{FAST}. The \ac{SKA} will also act as a \ac{PTA} when it comes online.

As can be seen in figure~\ref{Fi:GW_sens}, there is a gap in observing frequencies over the range $\sim10^{-7.5}$ -- $10^{-5.5}$ Hz. A variety of methods have been suggested to probe this regime (e.g., \cite{kachelriess2024,Blas_2022,martens2023}), but none have yet been funded.

\begin{table}[htbp]
\centering
\resizebox{\linewidth}{!}{
\begin{tabular}{m{2.5cm}|m{2.2cm}m{1.5cm}m{2cm}m{1.6cm}m{1.2cm}}
\hline
Experiment&Freq. Range $\log_{10}(f)$&Arm Length (km)&Operational&$T_\textrm{obs}$ (yr)&Ref(s).\\
\hline
\multicolumn{6}{c}{Ground-Based Experiments:} \\
\hline
LVK (collab.)&$[{1.3},{3.6}]$&$3$--$4$&2015/19/20$^\dag$ &5/5/5$^\dag$&\cite{LVK_doc}\\
CE&$[{0.7},{3.5}]$&$40$&2035--40*&5&\cite{evans2021}\\
ET&$[{1},{3.3}]$&$20$&2035*&5&\cite{ET_science,ET_doc}\\
\hline
\multicolumn{6}{c}{Space-Based Experiments:} \\
\hline
LISA&$[{-5},{-1}]$&$2.5\times 10^{6}$&2037*&4&\cite{LISA_doc}\\
DECIGO&$[{-3},{1}]$&1,000&2030s /40s*&3&\cite{DECIGO_doc}\\
BBO&$[{-2},{1}]$&50,000&TBA*&10&\cite{BBO_doc,Yagi_2011}\\
\hline
\multicolumn{6}{c}{Pulsar Timing Arrays:} \\
\hline
NANOGrav&$[{-8.75},{-7.5}]$&N/A&2004*&15+&\cite{NG_doc}\\
IPTA (collab.)&$[{-9},{-7.5}]$&N/A&2016$^\ddagger$&12.5+&\cite{IPTA_doc}\\
SKA&$[{-9},{-7}]$&N/A&2028*&20&\cite{SKA_doc}\\
\hline
\end{tabular}
}
\caption{Relevant properties of the \ac{GW} experiments considered in this paper.\\ \noindent\footnotesize{$^*$ Expected operational dates. $^\dag$ Operational phases 3/4/5, respectively. $^\ddagger$ First data release.}\label{Ta:GWexps}}
\end{table}

\subsection{Astrophysical Foreground---Binary Systems}\label{SSec:binaries}
For this work we consider the \ac{GW} spectra of inspiralling binary populations as a foreground contaminant for those of cosmological phenomena. Different binary masses contribute at different frequencies. At \ac{PTA} frequencies (nHz) the relevant binary systems are \ac{SMBHBs}, while at space-based interferometer frequencies \ac{WD} and \ac{SOBH} binaries contribute, and binaries containing \acp{SOBH} and \acp{NS} contribute at ground-based interferometer frequencies, as can be seen in figure~\ref{Fi:GW_sens}.

The gravitational waves we observe from a single binary at a given frequency $f_\textrm{GW}$ depend only on the distance of the galaxy $D$, and the chirp mass of the system $\mathcal{M}$:

\begin{equation}
	h_{\textrm{GW}}(f_\textrm{GW}) = \frac{2(G\mathcal{M})^{5/3}\pi ^{2/3}f_\textrm{GW}^{-2/3} }{c^4 D},
\end{equation}

Moving from $h$ to $\Omega$ using equation~\ref{eq:omega_h}, the power spectrum for a single binary is therefore described by a power law:
\begin{equation}\label{eq:FG_PL}
    \Omega_\textrm{GW}\propto \frac{\mathcal{M}^{10/3}}{D^2} \times f^{2/3}.
\end{equation}
The \ac{GW} spectrum of a population of randomly distributed binaries hence retains the same overall spectral shape as that of an individual binary.

There is no a priori lower limit on the frequencies emitted by each population, but a rough upper limit can be found by assuming a regime where both objects in a binary are spinless and of equal mass---that is, ignoring \ac{EMRI}\footnote{Eccentricity in binaries boosts \acp{GW} at higher frequencies \cite{Kelley}, while extreme eccentricity ($\gtrsim 0.9$) results in attenuation of \ac{GW} strength across the frequency range \cite{Enoki}.} Taking Kepler's third law, $\omega^2 \propto M^2/a^3$, where $a$ is the orbital separation, we can input the minimum orbital separation to establish the maximum possible orbital frequency. For compact objects (\acp{NS} and \acp{WD}) $a^\textrm{min}$ is simply twice their radius, and for black holes the Schwarzschild radius can be substituted. Finally noting that the frequency of the associated \acp{GW} is twice the orbital frequency, we are left with
\begin{alignat}{3}
     f_\textrm{GW}^{\textrm{max}}  && \ \propto \  \omega^{\textrm{max}} \ \propto \ &\frac{M^{1/2}}{R^{3/2}} \ &&\textrm{ (compact objects)},\\
    f_\textrm{GW}^{\textrm{max}} && \ \propto \ \omega^{\textrm{max}}  \ \propto \ &\frac{1}{M} \ &&\textrm{ (black holes)}.
\end{alignat}

The combined power spectrum for a population of inspiralling binaries is a superposition of each individual binary, and as such can be calculated from a population mass spectrum.

\subsection{Possible SGWB Sources}

\begin{figure}
\centering
\includegraphics[width=\linewidth]{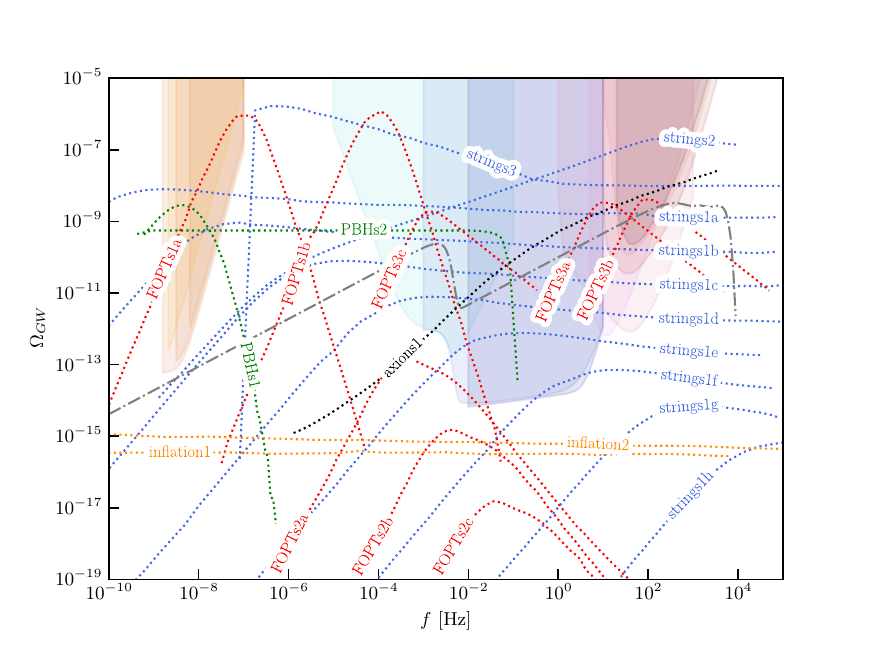}
\caption{Example spectra from a range of cosmological sources, detailed in Appendix \ref{A:fiducial_models}. These are intended to demonstrate the range of shapes and amplitudes of possible cosmological spectra, and are not intended as suggestions of anything specific that we expect to observe. As in figure~\ref{Fi:GW_sens}, the grey dash-dot line shows the total instantaneous foreground signal.}
\label{Fig:example_spectra}
\end{figure}

A variety of cosmological phenomena which might contribute to the \ac{SGWB} has been suggested. This includes inflation, \acp{PBH}, cosmic strings, \acp{FOPT}, and exotic phenomena including cosmic superstrings and axions/axion-like particles. The spectra of these cosmological sources can be modelled as single or broken power laws, as can bee seen in figure~\ref{Fig:example_spectra}. The precise connections between the physical parameters of a phenomenon and the spectral parameters of the power laws are complex, but there are correlations of the amplitude, peak and slope(s) of the power law with the physical parameters of each phenomenon. 

In all cases, evaluating the sensitivity of experiment to the amplitude and slope of a spectrum enables us to trace back sensitivities to physical parameters, either analytically or statistically.

\subsubsection{Inflation}
The same amplification of quantum fluctuations during a period of accelerating inflationary expansion which generates anisotropic structure across the universe via scalar fluctuations also generates \acp{GW} through the magnification of tensor perturbations. These tensor perturbations then contribute a background of \acp{GW} to the \ac{SGWB} \cite{Guzzetti}.

The resulting \ac{GW} spectrum is expected to be broad across frequencies, though the exact shape depends on the precise model of inflation involved.

We describe the primordial spectrum of tensor perturbations which arises from inflation as a power-law:
\begin{equation}\label{Eq:inflation}
	P_t(k) = r A_s(k_*) \left( \frac{k}{k_*} \right) ^{n_T},
\end{equation}
where $r = P_t/P_s$ is the tensor-to-scalar ratio ($P_s$ is the scalar power spectrum), $A_s$ is the amplitude of primordial \textit{scalar} perturbations at a given scale $k = k_*$, and $n_T$ is the tensor spectral index. Note that $n_T < 0 $ gives a red spectrum, $n_T>0$ a blue spectrum, and $n_T=0$ a scale-invariant spectrum.

Different theories predict their own values of $r$ and $n_T$, both of which we have yet to properly constrain. Data from the most recent Planck release \cite{Planck_2018} gives us an upper bound on  $r$, which has been derived from the combination of the Planck results and the BICEP2/Keck 2015 data to give an upper limit of $r < 0.044$ \cite{Tristram_2021}.\footnote{Note that this calculation was done under the assumption of \ac{SFSR} inflation, which requires the consistency relation $n_T = -r/8$. As such, it is not fully generalisable, and in some cases the previous constraint $r < 0.06$ is more appropriate.}

To observe these primordial tensor fluctuations as red-shifted gravitational waves today, we measure the present gravitational wave density $\Omega_\textrm{GW}$. Assuming the standard radiation, matter, then dark energy dominated eras, and at \ac{PTA} frequency, this can be approximated as \cite{Vagnozzi_2020}
\begin{equation}
	\Omega_\textrm{GW} \propto \frac{\Omega^2_m r A_s}{H^2_0 k^2_\textrm{eq}}\left(\frac{k}{k_*}\right) ^{n_T} \propto \left( \frac{f}{f_\textrm{ref}} \right) ^{n_T},
\end{equation}
where $k_\textrm{eq} \approx 0.073\Omega_m h^2$ Mpc$^{-1}$ is the matter-equality wavenumber. Hence the \ac{GW} power scales with $r$, with its shape dictated by $n_T$.
\subsubsection{Primordial Black Holes}

\acp{PBH} are formed from overdensities in the early universe, where scalar perturbations give a density contrast over some critical value $\delta \rho/\rho = \delta_c$.

At linear order, the tensor and scalar modes of a perturbation are independent. However, at higher orders, this is not true---the same scalar perturbations which lead to \acp{PBH} also induce \acp{GW}. It should be noted that the \acp{PBH} themselves are not generating the \acp{GW}; rather both phenomena are consequences of the same initial scalar perturbations.

In the same way that \acp{PBH} form when the relevant spherical region reenters the horizon, so too do \acp{GW} only begin to propagate when their wavelength reenters the horizon. Thus, we can explicitly link the frequency of the \acp{GW} not only to the time of their formation via the horizon size, but also to the mass of the coeval \acp{PBH}. The present day frequency is then approximately 
\begin{equation}
    f_\textrm{GW} \sim 10^{-9} \textrm{ Hz} \left( \frac{M_\textrm{PBH}}{30 M_\odot} \right)^{-1/2}.
\end{equation}

Increasing the mass of the \acp{PBH} shifts the peak of the \ac{GW} spectrum to lower frequencies. The amplitude of the spectrum is increased by increasing $f_\textrm{PBH}$, the proportion of dark matter attributed to \acp{PBH}.

We can also constrain \acp{PBH} by considering, e.g., Hawking radiation, positron interactions, micro-lensing, and distortions to the \ac{CMB} through Silk damping \cite{atal_2020}. The constraints on the abundance of \acp{PBH} at various mass ranges from a variety of experiments are explored in \cite{Kuroyanagi_2021}.

\subsubsection{Cosmic Strings}\label{SSS:cosmic_strings}
Cosmic strings are theoretical linear topological defects produced in the early universe. They form kinks and cusps as they vibrate, leading to the production of separate loops, which produce \acp{GW} as they decay. A population of cosmic strings would then generate a \ac{SGWB} \cite{Christensen_2018}.  These cosmic strings would be produced following a symmetry breaking event \cite{Kibble_1976} at the end of inflation \cite{Sakellariadou_2009}. Strings are parameterized by their tension $G\mu$, where $\mu$ is their mass per unit length. In units where $c=1$, $G\mu$ is dimensionless.

The peak amplitude of the resulting \ac{GW} power spectrum is proportional to $\Gamma \times\left(G\mu\right)^2$, where $\Gamma$ is the number density of loops \cite{strings1}. Cosmic string spectra are almost completely flat across the whole \ac{GW} experiment frequency range---see figure~\ref{Fig:example_spectra}.

We have upper bounds on the string tension from null results at various frequencies. The LIGO/Virgo restricts the tension to $G\mu< 10^{-8}$ \cite{Aasi_2014}, and the CMB requires $G\mu< 10^{-7}$ \cite{2014}. A combination of these constraints with additional factors from baryonic acoustic oscillations and gravitational lensing requires that $G\mu < 4\times 10^{-9}$ \cite{henrotversille:in2p3-01107320}. Reference \cite{Abbott_2018} considers various models of cosmic strings which satisfy these constraints.

\subsubsection{First-Order Phase Transitions}
The $\Lambda$CDM model of cosmology predicts second-order cosmological phase transitions, which shift the universe from one state to the next in a continuous way. However, \acp{FOPT}, which have discontinuous change, proceed via bubble nucleation---bubbles of the new state nucleate stochastically across the volume, expanding and merging until the phase transition is complete \cite{mazumdar_2019}. Only this type of transition generates gravitational waves. An observation of gravitational waves from \acp{FOPT} would tell us both about the history of the universe, but also requires new physics \cite{linde_1983}. \acp{FOPT} for a range of cosmological phase transitions have been proposed as generators of the \ac{SGWB}, including the electro-weak phase transition.

\acp{FOPT} can be characterised by four parameters: their strength $\alpha$, the rate of bubble nucleation $\Gamma$, the temperature when bubble nucleation begins $T_n$, and the speed of the expansion of the bubbles (the wall-velocity) $v_{\textrm{wall}}$. There is not a simple analytical model to describe the \ac{GW} spectrum from a \ac{FOPT}, nor are all \acp{FOPT} similar enough to have a single model. Depending on the four parameters, the transition (and hence also \ac{GW} spectrum) is qualitatively different. 

There are three elements of \acp{FOPT} which generate gravitational waves: bubble collisions, sound waves, and turbulence \cite{huang_zhang_2019}. \ac{GW} spectra from \acp{FOPT} have to be simulated, since it is impossible to derive their spectra analytically (although semi-analytic treatments have been carried out alongside simulations). The resulting spectra depend on $\alpha$, $\Gamma$, $T_n$, and $v_\textrm{wall}$, and strongly depend on the dominant effect---bubble collisions dominate only where there is very strong supercooling; sound waves for weak phase transitions, and turbulence for strong.

Where bubble collisions dominate, the spectrum is largely dictated by the average separation of bubbles $r_*$ (which can be derived from $\Gamma$ and $v_\textrm{wall}$); here an increase in $r_*$ increases the amplitude of the spectrum. Where sound waves dominate, increasing $r_*$ or $\alpha$ increases the amplitude and shifts the peak to lower frequencies; increasing $T_n$ increases the peak frequency; and increasing $v_\textrm{wall}$ slightly decreases the amplitude while modifying the shape of the spectrum slightly. Describing spectra from turbulence qualitatively is more complicated, but primarily the amplitude and frequency of the peak increase as $\alpha$ increases.

\subsubsection{Exotic Phenomena}
Other phenomena which might contribute to a \ac{SGWB} have been proposed, including cosmic superstrings and axions/axion-like particles (ALPs).

Cosmic superstrings, generated by the expansion of fundamental superstrings via inflation, would generate \acp{GW} in the same manner as cosmic strings, discussed in section~\ref{SSS:cosmic_strings} \cite{Damour_2005,Copeland_2010}, and as such would have spectra of the same shape. Unlike a detection of cosmic strings, a detection of superstrings would be evidence for string theory \cite{Kuroyanagi_2012}. Observational upper limits on the SGWB place bounds on the superstring tension, for various reconnection probabilities $p$: $G\mu<10^{-12}$ for $p\approx10^{-3}$, $G\mu \approx 10^{-10}$ for $p\approx10^{-1}$, and $G\mu \approx 10^{-8}$ for $p\approx1$ \cite{Siemens_2007}.

Axions are theorized low mass ($\sim10^{-5}$--$10^{-3}$ eV) spin- and chargeless particles, which follow from the Peccei-Quinn solution to the strong CP problem \cite{axions,axionsB}, which could also act as cold dark matter \cite{Duffy_2009}. Axion-like particles are similar in mass, and largely arise with the introduction of string theory to the standard model \cite{ringwald2014}.
The oscillations of an ALP field in the early universe would generate a gravitational wave background at some level \cite{Salehian_2021}, with specifics of the spectral shape depending on the precise model.

\section{Constraints on the SGWB}\label{Se:Constraints}

\subsection{Observations from PTAs}
\label{SSec:PTA_obs}
In 2022, NANOGrav's 15 year data releases (hereafter `NANOGrav15') gave the first confirmed observation of the \ac{SGWB} \cite{NG15_pulsars,NG15_gwb}. This observation has since been corroborated by other \acp{PTA}: \ac{EPTA} (including pulsars from \ac{InPTA}) \cite{EPTA_gwb}, \ac{PPTA} \cite{PPTA_gwb}, and the \ac{CPTA} \cite{CPTA_gwb}. A combined analysis in the upcoming \ac{IPTA} third data release will provide more sensitive constraints on the observations \cite{NG15_pulsars}.

PTA data are typically released in terms of the constraints on a power-law spectrum with a reference frequency of $f_\textrm{yr} = \textrm{yr}^{-1}$:
\begin{equation}
	h_c(f) = A_\textrm{CP}\left( \frac{f}{f_\textrm{yr}}\right)^\alpha
\end{equation}
and the cross-power spectral density of the timing deviations between two pulsars: 
\begin{equation}
	S_{ab}(f) = \Gamma_{ab} \frac{A^2_\textrm{CP}}{12\pi^2} f_\textrm{yr}^{-3}\left( \frac{f}{f_\textrm{yr}}\right)^{-\gamma_\textrm{CP}}
\end{equation}
where $\Gamma_{ab}$ is the Hellings-Down correlation coefficient between the two pulsars \cite{Hellings}. In the case where pulsar timing residuals are due entirely to a common \ac{GW} process, the two equations are linked such that $\alpha = (3-\gamma_\textrm{CP})/2$, and the fit of results to the Hellings-Down curve is therefore a test for the presence of the \ac{SGWB}. We can then obtain the gravitational wave energy density $\Omega$ using eq.~\ref{eq:omega_h}.

The \ac{PTA} analysis results in table~\ref{tab:results} are given in both joint $A_\textrm{GWB}$-$\gamma$ posteriors, and $A_\textrm{GWB}$ posteriors with a specified  $\gamma$ for \ac{SMBHBs} (the dominant astrophysical foreground population at PTA frequencies)---from eq.~\ref{eq:FG_PL}, the conversion between $\alpha$ and $\gamma$ above gives a value $\gamma_\textrm{SMBHBs}=13/3$.

\begin{table}
    \centering
    \begin{tabular}{c|rrrrrr}
        \hline
         & \textbf{NANOGrav} & \multicolumn{2}{c}{\textbf{EPTA}} & \multicolumn{2}{c}{\textbf{PPTA}} & \textbf{CPTA}\\
         \hline
        $\log_{10}A $ & $-14.61${\raisebox{0.5ex}{\tiny$^{+0.12}_{-0.11}$}} & \multicolumn{2}{c}{${-14.61}${\raisebox{0.5ex}{\tiny$^{{+0.11}}_{{-0.15}}$}}} & \multicolumn{2}{c}{$-14.69${\raisebox{0.5ex}{\tiny$^{+0.05}_{-0.05}$}}} & $-14.7${\raisebox{0.5ex}{\tiny$^{+0.9}_{-1.9}$}}\\
        \hline
        $\gamma$ & $\mathbf{3.2}${\raisebox{0.5ex}{\tiny$\mathbf{^{+0.6}_{-0.6}}$}} & $\mathbf{4.19}${\raisebox{0.5ex}{\tiny$\mathbf{^{+0.73}_{-0.63}}$}} & 
        $\mathbf{2.71}${\raisebox{0.5ex}{\tiny$\mathbf{^{+1.18}_{-0.73}}$}} & $\mathbf{3.87} $ & 
        $\mathbf{4.02} $ & $[0,6.6]$\\
        $\alpha$ & $-0.1${\raisebox{0.5ex}{\tiny$^{+0.3}_{-0.3}$}} & $-0.60${\raisebox{0.5ex}{\tiny$^{+0.32}_{-0.36}$}} & $0.15${\raisebox{0.5ex}{\tiny$^{+0.36}_{-0.60}$}} & $-0.44 $ & $-0.51 $ & $\mathbf{[-1.8,1.5]}$\\
        $\log_{10}A$ & $-14.19${\raisebox{0.5ex}{\tiny$^{+0.22}_{-0.24}$}} & $-14.54${\raisebox{0.5ex}{\tiny$^{+0.28}_{-0.41}$}} & $-13.94${\raisebox{0.5ex}{\tiny$^{+0.23}_{-0.48}$}} & $-14.5 $ & $-14.56 $ & $-14.4${\raisebox{0.5ex}{\tiny$^{+1.0}_{-2.8}$}}\\
        \hline
    \end{tabular}
    \caption{Analysis results from each \ac{PTA} given at the 95\% CL (NANOGrav \cite{NG15_pulsars,NG15_gwb}, \ac{EPTA} (including pulsars from \ac{InPTA}) \cite{EPTA_gwb}, \ac{PPTA} \cite{PPTA_gwb}, \ac{CPTA} \cite{CPTA_gwb}). Values in the upper section consider a fixed value of $\gamma=13/3$ ($\alpha = -2/3)$, i.e., a population of \ac{SMBHBs}. Values in the lower section are joint $A$-$\gamma$ and $A$-$\alpha$ posteriors. Values in bold indicate which of $\gamma$ and $\alpha$ the original joint posteriors were given in---conversions have been made for convenience of comparison. Where no errors are listed, errors were not provided in the original references. Ranges given in square brackets indicate uniform sampling across the range (insufficient observing time for constraints).}
    \label{tab:results}
\end{table}

\subsection{Upper Bounds from Ground-Based Detectors SGWB}\label{SS:upperbounds}
The most up-to-date upper bounds from operational ground-based detectors comes from the \ac{LVK}-O3 data release \cite{LVK_gwb}. Table~\ref{tab:LVK_results} gives the upper limits on the \ac{SGWB} signal at a reference frequency of 25 Hz from that work, for three values for the power-law index: $m=0$ (a flat spectrum, corresponding to, e.g., cosmic strings or \ac{SFSR} inflationary models), $2/3$ (compact binary inspiralling population), and 3 (historically used value, e.g., some supernova, see \cite{LVK_gwb} and references therein). Note that $m$ is named $\alpha$ in the original \ac{LVK} analyses; it has been changed here to avoid confusion with $\alpha$ used in Section~\ref{SSec:PTA_obs}, and for consistency with our results later.

\begin{table}
    \centering
    \begin{tabular}{c|cc}
        \hline
         & \multicolumn{2}{c}{$\Omega_\textrm{GW}\times10^{-8}$} \\
         \hline
          $m$ & Uniform prior & Log-uniform prior \\
           \hline
        0 & 1.7 & 3.5\\
        $2/3$ & 1.2 &3.0 \\
        3 & 0.13 & 0.51 \\
        \hline
       
    \end{tabular}
    \caption{Current upper limits from \ac{LVK}-O3, at the $95\%$ CL \cite{LVK_gwb}. Results are given for uniform and log-uniform priors on $\Omega_\textrm{GW}(f_\textrm{ref} = 25 \textrm{ Hz})$, where the former is more conservative and the latter is more sensitive to weak signals.}
    \label{tab:LVK_results}
\end{table}

\subsection{Constraints from the CMB}
Current \ac{CMB} observations also provide an upper limit on the \ac{SGWB} at frequencies even lower than \acp{PTA}.

In \ac{CMB} temperature observations, present-day gravitational waves would induce an additional quadrupole in the \ac{CMB}; historic gravitational waves propagating at the formation of the \ac{CMB} would have caused damping on the smallest angular scales. 

Any gravitational wave background already propagating by the time of \ac{CMB} generation (recombination) would introduce $B$ modes into an otherwise $B$-mode free \ac{CMB} \cite{doi:10.1146/annurev-astro-081915-023433}. Separating out $B$ modes from gravitational lensing by objects between us and the \ac{CMB} (dust within the galaxy, but also massive objects further away) should leave us with observable $B$ modes from gravitational waves; we expect to detect these at angular scales of $\sim 0.1^{\degree}$--$1^{\degree}$ \cite{doi:10.1146/annurev-astro-081915-023433}. 

Since we have yet to detect any of these effects, the temperature, $B$-mode polarization and lensing from Planck and BICEP/Keck place an upper limit of $\Omega_\textrm{GW} \lesssim 10^{-12} ({k}/{0.1\textrm{ Mpc}^{-1}})^3$ for $k \gtrsim 0.1 \textrm{ Mpc}^{-1}$ (corresponding to frequencies between roughly $10^{-16}$--$10^{-14}$) \cite{Namikawa:2019tax}.

\section{Methodology}\label{Se:Methodology}
Our overall goal is to determine the constraints from current and future GW experiments, individually and in combination, on different sources of the SGWB. We model SGWB sources as single power laws, in the presence of astrophysical foregrounds.

\subsection{Modelling the SGWB}\label{SSec:modelling}
The strength of the \ac{SGWB} can be described by its energy density 
\begin{equation}
   \Omega_\textrm{GW}\equiv \frac{1}{\rho_c}\frac{\textrm{d}\rho_\textrm{GW}}{\textrm{d}\ln f},
\end{equation}
where $\rho_c$ is the critical density. The energy density can be converted to other measures:

\begin{equation}\label{eq:Omega-S}
    \Omega_{\textrm{GW}}(f) = \frac{2 \pi^2}{3H^2_0} f^3 S_h(f) = \frac{2 \pi^2}{3H^2_0}f^2[h_c(f)]^2 = \frac{8 \pi^2}{3H^2_0} f^4 |\Tilde{h}(f)|^2,
\end{equation}
where $S_h$ is the (one-sided) \ac{PSD}, $h_c$ is the characteristic strain, and $\Tilde{h}=\sqrt{|\Tilde{h}_+|^2 + |\Tilde{h}_\times|^2}$ is its Fourier transform. With the exception of $S_h$ and $\Tilde{h}$, which both have units of Hz$^{-1}$, all of these measures are dimensionless. 

\subsubsection{Instrumental Noise} \label{SSSec:Ins_Noise}
The noise spectrum of an instrument can be defined similarly. The \ac{PSD} of a single detector $S_\textrm{ins}(f)$ is 
\begin{equation}
    S_\textrm{ins}(f) \equiv \frac{N(f)}{\mathcal{R}(f)} ,
\end{equation}
where $N(f)$ is the instrumental noise and the expression defines $\mathcal{R}(f)$, the detector response function, such that $S$ is the \ac{PSD} required to produce a signal equal to the instrumental noise. A value of $\mathcal{R}(f)=1$ denotes a perfect experiment. $S_\textrm{ins}^\textrm{eff} = S_\textrm{ins} / \sqrt{n_\textrm{det}}$ is the effective noise \ac{PSD} of a network of $n_\textrm{det}$ detectors, and $\Omega_\textrm{ins}$ is then $S_\textrm{ins}^\textrm{eff}$ expressed in terms of an energy density.\footnote{Note that some papers work with $\Omega_\textrm{ins}(S_\textrm{ins})$ rather than $\Omega_\textrm{ins}(S_\textrm{ins}^\textrm{eff})$, and care should be taken with factors of $n_\textrm{det}$.}

\begin{equation}\label{eq:ins_noise}
    H^2_0 \Omega_{\textrm{ins}}(f) = \frac{2 \pi^2}{3} f^3 S_{\textrm{ins}}(f) = \frac{2 \pi^2}{3}f^2[h_{\textrm{ins}}(f)]^2 = \frac{8 \pi^2}{3} f^4 |\Tilde{n}(f)|^2 ,
\end{equation}
where $S_\textrm{ins}$ is the (one-sided) \ac{PSD}, $h_\textrm{ins}$ is the noise strain, and $\Tilde{n}$ is its Fourier transform.

\subsubsection{Foregrounds}\label{SSS:foregrounds}
The astrophysical foregrounds at different frequencies are generated by inspiralling binary populations of different masses (see discussion in section~\ref{SSec:binaries}). Where it is possible to model both the total and unresolved foreground spectra (that is, the entire spectra and what remains after resolution of individual binaries), results are shown with both. Results are also shown without any inclusion of astrophysical foregrounds, to represent ideal, expected, and worse observations.

We model the different populations of binaries such that their energy densities are described by the function
\begin{equation}\label{eq:FG_tanh}
\Omega_\textrm{fg} = A f ^{2/3} \left[ \ \exp \left( - (f/f_\textrm{ref} )^4 \right) + D \ \sin^2 \left( \pi f / f_\textrm{ref} \right)\right] \times \left[ 1 + \tanh \left( E \left( f_* - f\right) \right) \right] ,
\end{equation}
where a set of coefficients $A$, $D$, $E$,$f_*$, and $f_\textrm{ref}$ for each population is given in appendix~\ref{A:foreground}, found with a least-squares fitting~\ref{eq:FG_tanh} to the models detailed in appendix~\ref{A:foreground}. Equation~\ref{eq:FG_tanh} is an adjustment of a similar \ac{WD} binary model \cite{Cornish_2017}, to provide a more general fit across different populations.

A more comprehensive spectral shape can be constructed with the inclusion of merger and ringdown spectra, which is detailed in \cite{Ajith}. This affects the shape of the spectrum to the right of the merger frequency (i.e., the cutoff for inspiral signals) an approximation of which frequency is given in section~\ref{SSec:binaries}). For many of the experiments considered here this behaviour is below the relevant sensitivities, but see \cite{Bi_2023,steinle2023implicationspulsartimingarray} for more detail on extended spectra of \ac{SMBHBs} which might be visible in the \ac{LISA} band. 

The total observed power is therefore given by 
\begin{equation}
\begin{split}
\Omega_\textrm{total} &= \Omega_\textrm{GW}+\Omega_\textrm{noise}\\
&= \Omega_\textrm{GW}+\Omega_\textrm{ins}+\Omega_\textrm{fg} .
\end{split}
\end{equation}

In the case where multiple sources contribute to the \ac{SGWB} at a given frequency, $\Omega_\textrm{GW}$ can be taken to refer solely to a source of interest, with the contributions of the remaining source(s) contained within the overall $\Omega_\textrm{noise}$, which is taken to be known and fixed. Alternatively, $\Omega_\textrm{GW}$ can be taken as the combined power of the sources, in which case we must fit for the parameters of each such source, but this comes at the cost of lower sensitivity to, or even degeneracies between, those parameters. This is explored further in our upcoming paper \cite{paper2}.

\subsection{Fisher Matrices and Sensitivities to the SGWB}\label{subs:Fisher}
Fisher matrices can be used to learn about the sensitivities of an experiment to specified parameters. These parameters could be theoretical/`physical' (where the parameters link the theory being investigated directly to the signal being observed) or `phenomenological' (where the parameters simply characterize the spectral shape of the signal).

The Fisher matrix can be written such that its elements are the expectation values of the second derivatives of the log-likelihood with respect to some set of parameters $\mathbf{\theta}$, 
\begin{equation}
    F_{ij}= \left< \frac{\partial ^2 l_G}{\partial \theta_i \partial \theta_j} \right>.
\end{equation}
 The inverse matrix $C=F^{-1}$ is then a useful approximation to the covariance matrix.

In line with extant \ac{GW} experiment practices, we consider logarithmically binned frequency observations: the width of each bin, defined by its (logarithmically) central frequency $f_b$, is given by $\delta f_b$. The Fisher matrix used in this work, assuming uncorrelated Gaussian likelihoods in the set of bins $b=1,\ldots,N_b$, and is derived in \cite{Gowling_2021}, is
\begin{equation}\label{EQ:Fisher_matrix}
    F_{ij} = \sum^{N_b}_{b=1} \frac{1}{\sigma_{b,\Omega}^2}\frac{\partial\Omega_\textrm{GW}}{\partial\theta_i}\frac{\partial\Omega_\textrm{GW}}{\partial\theta_j},
\end{equation}
where $\sigma_{b,\Omega}^2$ is the variance of $\Omega_\textrm{GW}$ in bin $b$,\footnote{Note that this gives the variance when only one experiment is involved. Additional experiments which observe in the same frequency bin lead to a reduction in error. See section~\ref{SSec:multiple}.} given by
\begin{equation}\label{eq:sigmab2}
\frac{1}{\sigma_{b,\Omega}^2}=T_\textrm{obs}\frac{2\delta f_b}{\Omega_{\textrm{total}}^2}.
\end{equation}
Hence the variance depends not only on the total observation time of the experiment $T_\textrm{obs}$, the width of the frequency bins $\delta f_b$, and total number of frequency bins $N_b$, but also (as discussed in section~\ref{SSSec:nwt}) on $\Omega_\textrm{total}$---we cannot observe the 
\ac{GW} signal power in isolation but rather the total spectrum, polluted by the total noise contribution $\Omega_\textrm{noise}(f)$ including foregrounds and any other sources of \ac{GW} power in addition to instrumental noise (see section~\ref{SSS:foregrounds}).

The \ac{SNR} is a commonly used figure of merit for the detection of discrete sources, but is less useful when considering the \ac{SGWB}. The \ac{SNR} evolves with observing time $T_\textrm{obs}$, and can be written in terms of either 
$\Omega$ or $S$:
\begin{equation}\label{eq:SNR}
\begin{split}
\rho^2 = T_\textrm{obs} &\int_{f_\textrm{min}}^{f_\textrm{max}} \textrm{d}f \frac{\Omega_\textrm{GW}^2(f)}{\Omega_\textrm{noise}^2(f)} \\
= T_\textrm{obs} & \int_{f_\textrm{min}}^{f_\textrm{max}} \textrm{d}f \frac{{S_\textrm{GW}}^2(f)}{{S^\textrm{eff}_\textrm{noise}}^2(f)}.
\end{split}
\end{equation}
But when measuring a power spectrum, the error is determined by the total observed variance. This can be calculated from eq.~\ref{eq:sigmab2}, and notably does depends on $\Omega_\textrm{total}$ and not $\Omega_\textrm{noise}$. As such, the \ac{SNR} is not directly used in our results. 

\subsubsection{The Evolution of Noise with Time}\label{SSSec:nwt}

The evolution of noise with time leads us to plot multiple curves describing the instrumental noise contributions of each experiment: the `nominal' noise curve as given by the \acp{PSD} shown in appendix~\ref{A:noise}, and a \ac{nwt} curve which demonstrates the effective increase in sensitivity given by running an experiment for its predicted $T_\textrm{obs}$. We define these \ac{nwt} curves as 
\begin{equation}
    \Omega_\textrm{nwt}(f_b,T_\textrm{obs}) = \frac{\Omega_\textrm{noise}}{\sqrt{T_\textrm{obs}\delta f_b}} ; S_\textrm{nwt}(f_b,T_\textrm{obs}) = \frac{S^\textrm{eff}_\textrm{noise}}{\sqrt{T_\textrm{obs}\delta f_b}} .
\end{equation}
These \ac{nwt} sensitivity curves are intended as an alternative to the commonly used \acp{PLISC} (see \cite{Thrane_2013}), which show a curve above which the \ac{SNR} meets some threshold value. As discussed above, the \ac{SNR} is not a proxy for the variance when dealing with the \ac{SNR}, and as such \acp{PLISC} are not a useful tool to visualise the Fisher matrix-derived sensitivities of experiments. The nominal and \ac{nwt} sensitivity curves for each experiment considered are shown in figure~\ref{Fi:GW_sens}.

The observing frequency range of a given experiment, $\Delta f=f_\textrm{max}-f_\textrm{min}$, depends on the nature of the experiment. However the minimum \textit{possible} frequency also depends on $T_\textrm{obs}$, and is given by  $1/T_\textrm{obs}$, regardless of the frequencies eventually reachable by the experiment. Hence an observing time of 1 year gives a minimum possible frequency of $\sim 3 \times 10^{-8}$. For space- and ground-based experiments, this is well below the frequency range being probed. However for \acp{PTA}, which probe frequencies as low as $10^{-9}$, this means that their observing frequency range evolves with time. As such, the `nominal' noise curve for \acp{PTA} (\ac{IPTA}, \ac{SKA}) should not be considered to give an accurate description of the initial performance. Multiple \ac{nwt} curves are plotted for \ac{SKA} in figure~\ref{Fi:GW_sens} at different observing durations, to show its evolving frequency range.

\subsection{Power Law Sensitivities}\label{SS:power_law}

In this paper we test the sensitivity of each \ac{GW} experiment to the parameters for a single power law, defined by its amplitude $A$ and power-law index (or logarithmic slope) $m$. 

\begin{equation}\label{eq:powerlaw}
    \Omega = A\left(\frac{f}{f_0}\right)^m.
\end{equation}
A power law can be defined with a specified `pivot point' $f_0$, such that its amplitude $A$ at that point is constant but its power $m$ can vary. We can rewrite a power law as a straight line by moving to logarithmic parameters:
\begin{equation}
    Y = Y_0 + m (X-X_0),
\end{equation}
where $Y= \log_{10}(\Omega)$, $Y_0= \log_{10}(A)$, $X= \log_{10}(f)$, $X_0= \log_{10}(f_0)$. We can rewrite equation~\ref{EQ:Fisher_matrix} to obtain the Fisher matrix in $Y$ rather than $\Omega$:

\begin{equation}
    F_{ij} = \sum \frac{1}{\sigma_{b,Y}^2}\frac{\partial Y}{\partial\theta_i}\frac{\partial Y}{\partial \theta_j}, \textrm{     where     } \frac{1}{\sigma_{b,Y}^2}=\ln(10)^2T_\textrm{obs}\frac{2\delta f_b\Omega_\textrm{GW}^2}{\Omega_\textrm{total}^2}.
\end{equation}

This allows for a simple diagonalization of the Fisher and covariance matrices, de-correlating the parameters, by specifying the pivot frequency:
\begin{equation}
    X_0 = \sum_b\frac{1}{\sigma_b^2}X_b/\sum_b\frac{1}{\sigma_b^2},
\end{equation}
which gives:
\begin{equation}\label{eq:Fisher_final}
F =
\begin{bmatrix}
\sum_b\frac{1}{\sigma_b^2}X^2-\frac{\left(\sum_b\frac{1}{\sigma_b^2}X\right)^2}{\sum_b\frac{1}{\sigma_b^2}} & 0\\
0 & \sum_b\frac{1}{\sigma_b^2}
\end{bmatrix}, \textrm{     and     } C = 
\begin{bmatrix}
\sigma^2_m & 0\\
0 & \sigma^2_{Y_0}
\end{bmatrix}.
\end{equation}
The quantity $X_0$ is simply the weighted average of frequencies given their variances, and represents the point on the spectrum which can be best constrained.

\subsection{Combining Multiple Experiments}\label{SSec:multiple}
A benefit of the Fisher matrix formalism is the simplicity of considering the combined sensitivities of multiple experiments. This can be done simply by summing the Fisher matrices of each experiment:
\begin{equation}\label{eq:Fisher_comb}
    F_\textrm{combined}=F_\textrm{exp 1}+F_\textrm{exp 2}\ ; \ C_\textrm{combined} = F_{\textrm{combined}}^{-1} \ .
\end{equation}
$C_\textrm{combined}$ can then be used to find the improvement on $\sigma_Y^2$ and $\sigma_m^2$ of a combination of experiments, compared to each experiment individually. Note that equation~\ref{eq:Fisher_comb} requires a consistent fiducial \ac{GW} spectrum across the experiments being considered; here, that entails a single power law which stretches across the frequency range covered by the experiments and, as can be seen from equation~\ref{eq:Fisher_final}, a shared pivot frequency.

We can also calculate the new variance of a single frequency bin when a combination of experiments are considered, $\sigma_{b,Y,\textrm{ combined}}^2$:

\begin{equation}
    \sigma_{b,Y,\textrm{ combined}}^2 = 1/\left[1/\sigma_{b,Y,1}^2+ 1/\sigma_{b,Y,2}^2 + 1/\sigma_{b,Y,3}^2 \ldots \right]
\end{equation}

The reduction in variance of a single frequency bin of combined experiments compared to a given experiment 1, is then given by
\begin{equation}
    R = \frac{\sigma_{b,Y,\textrm{ combined}}^2}{\sigma_{b,Y,1}^2} = 1+ \sigma_{b,Y,1}^2/\sigma_{b,Y,2}^2 + \sigma_{b,Y,1}^2/\sigma_{b,Y,3}^2 \ldots
\end{equation}

\subsection{Adding an Observational Prior}\label{SS:prior}
With the Fisher matrix formalism, the inclusion of previous observations as a prior is identical to the inclusion of additional experiments---it can be done just through the addition of an extra Fisher matrix per section~\ref{SSec:multiple}, this time with the inverse of the observed error bars in the places of $1/\sigma_m^2$ and $1/\sigma_Y^2$. In this case, the fiducial model must match the mean observed values for $Y$ and $m$.

We can instead consider the case where the fiducial \ac{GW} spectrum under consideration deviates from the prior. By taking the distributions of both the fiducial model and the observed prior to be Gaussian and described by $\theta_p,\sigma_p$ and $\theta_d,\sigma_d$, respectively, we can find their combined distribution by taking their product. We aim to get an approximately Gaussian combined distribution. Dropping prefactors, the product of the two initial Gaussian distributions can be written as a Gaussian with mean and variance $\theta_*,\sigma^2_*$ using
\begin{equation}
    \frac{(\theta - \theta_p)^2}{\sigma_p^2} + \frac{(\theta - \theta_d)^2}{\sigma_d^2} = \frac{(\theta - \theta_*)^2}{\sigma_*^2} + \chi^2,
\end{equation}
where
\begin{equation}
\frac{1}{\sigma_*^2} = \frac{\sigma_p^2 + \sigma_d^2}{\sigma_p^2\sigma_d^2},
\qquad
\theta_* = \frac{\theta_p\sigma_d^2 + \theta_d\sigma_p^2}{\sigma_p^2 + \sigma_d^2},
\qquad
\chi^2 = \frac{(\theta_p -\theta_d)^2}{\sigma_p^2 + \sigma_d^2}.
\label{eq:sigma_star}
\end{equation}
This only depends on $\theta_*,\sigma^2_*$ when $\theta_p\approx \theta_d$. Hence the Fisher formalism can only be used to calculate the constraining power of experiments can only be made in this case, i.e., when considering a fiducial model which is close to the prior model. In this paper, we limit ourselves to cases where $|\theta_p - \theta_d|$ lies within the reported $95\%$ CL.

\section{Results}\label{Se:Results}

We consider first how well an individual experiment can constrain a single power law---what are its sensitivities to the slope and amplitude? Running a fiducial spectrum through the Fisher matrix gives us the minimum error on $Y_0$ and $m$ ($\sigma_{Y_0}$ and $\sigma_m$, respectively) for a given experiment. Examples are shown in figure~\ref{fig:results1}, where $\sigma_{Y_0}$ is shown as a shaded band above and below the fiducial spectrum, and $\sigma_m$ is shown as the shaded region between the upper and lower limits on slope, added to the limits on the amplitude. Both errors are centred on the pivot point, $f_0=10^{X_0}$, plotted as a star. This new method, displaying the decorrelated errors with the combined shaded region, does not fully capture the combination of errors but is a reasonable approximation.

We use 20 logarithmically spaced bins per experiment---changing the number of bins gives negligible impact on the results, provided that there are at least three bins. There is likewise negligible impact if linearly separated bins are used instead. Where bins are shared between multiple experiments (where we are evaluating bin-by-bin errors for combinations of experiments, for example) the total number of bins is chosen to approximately match the density of bins across frequency space in the individual analyses.

The fiducial models used in our results are chosen primarily to illustrate a generalisation of our findings, but are all taken from genuine physical spectra which have been put forward as \ac{SGWB} contributions from a range of phenomena, and can all be found in appendix \ref{A:fiducial_models}.

\subsection{Individual Experiments}\label{Results-indiv}
\begin{figure}[H]
\centering
\begin{subfigure}[b]{0.49\textwidth}
\centering
\includegraphics[width=\textwidth]{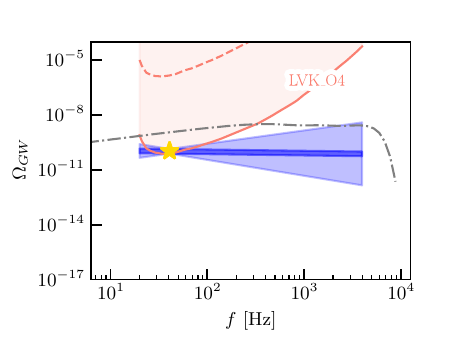}
\caption{}\label{subfig:LVK04}
\end{subfigure}
\hfill
\begin{subfigure}[b]{0.49\textwidth}
\centering
\includegraphics[width=\textwidth]{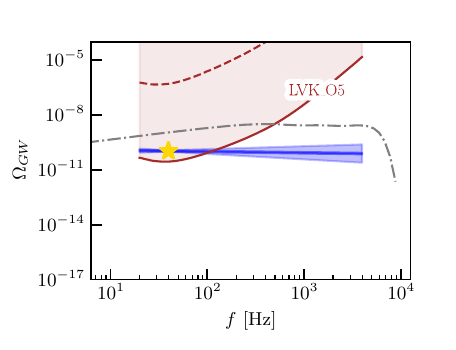}
\caption{}\label{subfig:LVK05}
\end{subfigure}
\hfill
\begin{subfigure}[b]{0.49\textwidth}
\centering
\includegraphics[width=\textwidth]{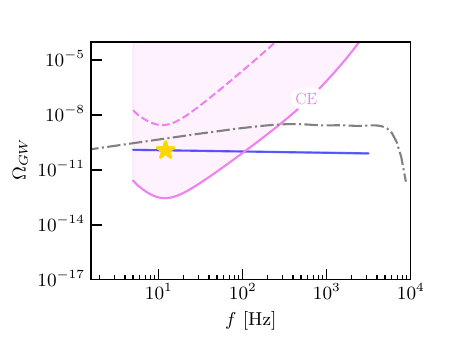}
\caption{}\label{subfig:CE}
\end{subfigure}
\begin{subfigure}[b]{0.49\textwidth}
\centering
\includegraphics[width=\textwidth]{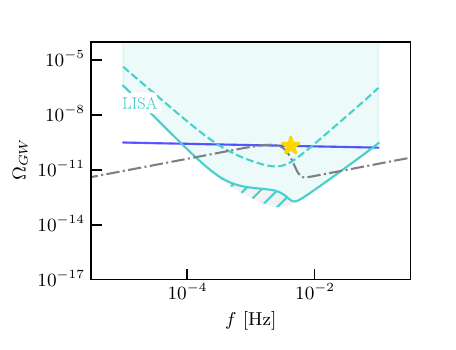}
\caption{}\label{subfig:LISA}
\end{subfigure}
\caption{The sensitivities of LVK O4, LVK O5, \ac{CE} and \ac{LISA} to (strings1d). Shaded sensitivity regions match those in figure~\ref{Fi:GW_sens}, and the grey dash-dot line is likewise the total instantaneous foreground. \\(a) LVK-O4: $f_0 = 40.8$ Hz, $\Omega_0 = 1.08 \times 10^{-10} \pm 1.17 \times 10^{-11}$, $m=-0.07 \pm 0.80$.\\ (b) LVK-O5: $f_0 = 39.7$ Hz, $\Omega_0 = 1.09 \times 10^{-10} \pm 1.00 \times 10^{-11}$, $m=-0.07 \pm 0.23$.\\ (c) CE: $f_0 = 12.2$ Hz, $\Omega_0 = 1.18 \times 10^{-10} \pm 1.32 \times 10^{-13}$, $m=-0.07 \pm 0.004$.\\ (d) LISA: $f_0 = 0.004$ Hz, $\Omega_0 = 2.06 \times 10^{-10} \pm 2.57 \times 10^{-13}$, $m=-0.07 \pm 0.002$.}
\label{fig:results1}
\end{figure}

Where the constraints on a spectrum are good, i.e., low $\sigma_{Y0}$ and $\sigma_m$, the shaded regions present as narrow bands along the fiducial line. Conversely, poor constraints generate wide `bowtie' shaped errors. Sensitivities of a selection of experiments to an example cosmic string spectrum (strings1d, \cite{strings1}) are shown in figure~\ref{fig:results1}.

The improving sensitivities of the three ground-based experiments (LVK-O4, LVK-O5, \ac{CE}) shown in figures \ref{subfig:LVK04}--\ref{subfig:CE}
can be seen both in the move of the shaded sensitivity regions to lower amplitudes, as well as in the shrinking of the `bowties' which represent the errors on the slope and amplitude---the `bowtie' collapses to a narrow band in cases where the fiducial spectrum can be very well constrained. Figure~\ref{subfig:LISA} then shows the sensitivity of \ac{LISA} to the same fiducial spectrum at lower frequencies, as a comparison. 

\begin{figure}[H]
\centering
\begin{subfigure}[t]{\textwidth}
\centering
\includegraphics[width=0.49\linewidth]{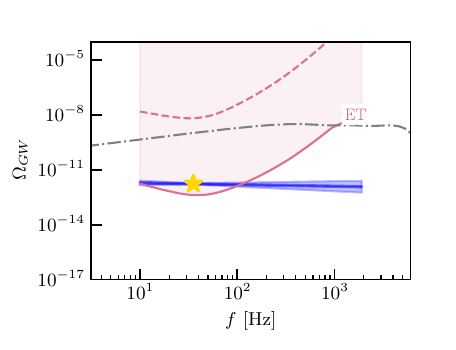}
\includegraphics[width=0.49\linewidth]{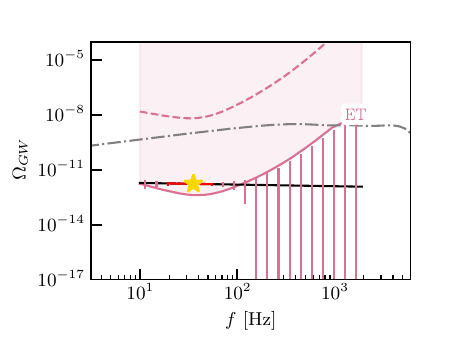}
\caption{}
\label{fig:ET}
\end{subfigure}
\begin{subfigure}[t]{\textwidth}
\centering
\includegraphics[width=0.49\linewidth]{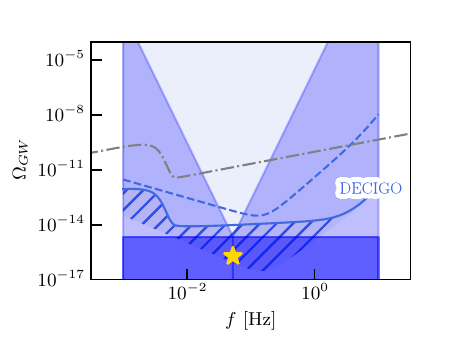}
\includegraphics[width=0.49\linewidth]{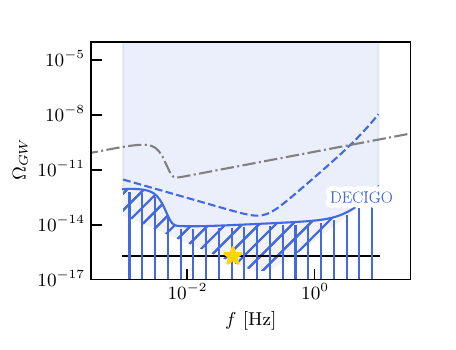}
\caption{}
\label{fig:BBO}
\end{subfigure}
\caption{A visualisation of the results given by $\sigma_{Y_0}$ and $\sigma_m$ (left plots) compared to $\sigma_{Y_0,b}$ (right). Where the fiducial model is plotted in red, $\sigma_{Y_0,b}/Y_b \leq 1 \%$, else it is plotted in black. \\
(a) ET (strings1d): $f_0 = 35.5$ Hz, $\Omega_0 = 1.73 \times 10^{-12} \pm 1.40 \times 10^{-13}$, $m=-0.09 \pm 0.168$. Range constrained to 1\%: 19.3--55.1 Hz.\\
(b) DECIGO (inflation1): $f_0 = 0.052$ Hz, $\Omega_0 = 1.91 \times 10^{-16} \pm 2.06 \times 10^{-15}$, $m=0 \pm 7.16$. Range constrained to 1\%: N/A.}
\label{fig:results2}
\end{figure}

As discussed in section~\ref{SSSec:nwt}, a fiducial spectrum which crosses above the $\Omega_\textrm{nwt}$ line should be observable. But it is not true that the entire observable frequency range of an experiment can be informative, as can be seen in the `well-constrained' frequency ranges in figure~\ref{fig:results2}---it can be seen that the width of the `well-constrained' frequency range is all-important in determining whether the \textit{shape} of the spectrum can be confidently assessed. 

Recall from section~\ref{SS:power_law} that the pivot point gives the frequency where the amplitude of the spectrum is best constrained---while contributions per frequency bin to $\sigma_m$ are constant, contributions to $\sigma_{Y_0}$ are not. That is, $\sigma_{Y_0,b}$  depends on the particular frequency bin $f_b$, and varies widely across the frequency ranges of each experiment. It is therefore useful to probe the actual frequency range over which the experiment can confidently assess the amplitude of a given frequency bin, since the Fisher matrix formalism assumes the fiducial model across the entire prescribed frequency range. This is particularly important in assessing the shape of an observed spectrum, as a region of well-constrained $\sigma_{Y_0,b}$ also gives the frequency range where the shape can be confidently determined. These ranges, specified as frequency bins where $\sigma_{Y_0,b}/Y_b \leq 1 \%$, are shown in red.

Figure~\ref{fig:results2} shows that there are some cases where `bowtie' results can be misleading with regards to the errors on the spectral shape: in  cases where the fiducial model dips below the sensitivity of an experiment, they can over- (figure~\ref{fig:ET}) or under-estimate (figure~\ref{fig:BBO}) the 1$\sigma$ regions of uncertainty. Figure~\ref{fig:ET} is a useful demonstration that the `bowtie' within the sensitivity region of an experiment is a good match to the shape of the individual error bars, but begins to fail outside of the sensitivity region. In this case the error bars show that there is a region of good constraint (shown in red) and a region where we can only establish an upper bound, while the `bowtie' remains misleadingly narrow across the whole frequency space. Where the spectrum lies entirely below the ultimate sensitivity of an experiment, an upper bound rather than observation follows: the shape of the experiment's sensitivity curve itself rules out not only a range of amplitudes, but also of slopes. In these situations, the `bowties' which are generated are often not the most useful result. Figure~\ref{fig:BBO} demonstrates that the $\sigma_{Y_0,b}$ sketch out a more useful area than $\sigma_{Y_0}$ and $\sigma_m$ when evaluating the sensitivity of \ac{DECIGO} to a \ac{SFSR} model of inflation. 

\subsection{Multiple Experiments}
We next evaluate the improvements to the constraints on slope and amplitude by using multiple experiments in conjunction, and also the regime over which we can be confident about the shape of the spectrum being observed. We divide combinations of experiments into two categories: where the experiments overlap in frequency space, and where they are distinct---i.e., they lie on either side of a gap in observing frequency.

\subsubsection{Overlapping Experiments}\label{Results-overl}
Within broad bands at low (pulsar timing) and high (ground-based) frequencies, upcoming experiments overlap in frequency space with the current experiments, to provide improved sensitivities. This can be seen in figure~\ref{Fi:GW_sens}, where the sensitivities progress to lower amplitudes moving from \ac{IPTA} $\rightarrow$ \ac{SKA}, and from \ac{LVK}-O3 $\rightarrow$ \ac{LVK}-O5 $\rightarrow$ \ac{CE} and \ac{ET}.

\begin{figure}[H]
\centering
\begin{subfigure}[b]{\textwidth}
\centering
\includegraphics[width=0.49\textwidth]{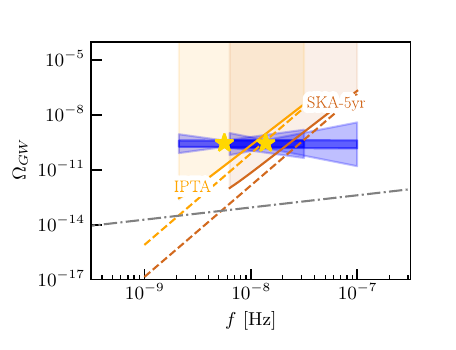}
\includegraphics[width=0.49\textwidth]{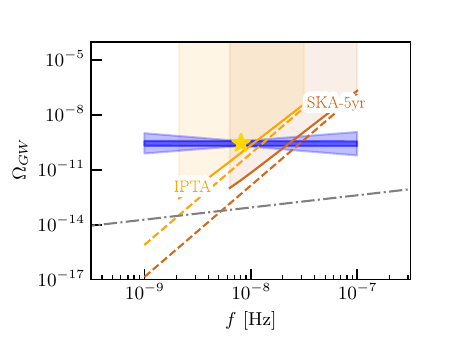}
\caption{}\label{subfig:overlappingA}
\end{subfigure}
\hfill
\begin{subfigure}[b]{\textwidth}
\centering
\includegraphics[width=0.49\textwidth]{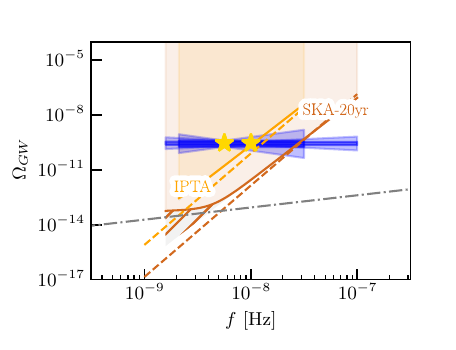}
\includegraphics[width=0.49\textwidth]{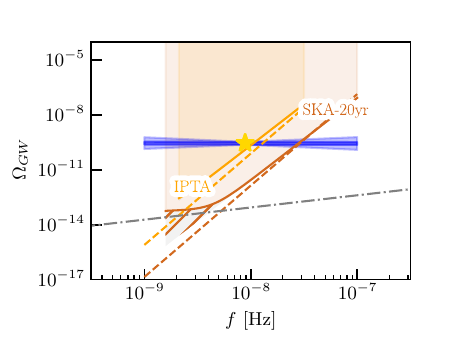}
\caption{}\label{subfig:overlappingB}
\end{subfigure}
\caption{Improvement in constraining power that IPTA confers to SKA where SKA has been observing for 5 years (top) and 20 years (bottom) to (PBHs2).\\
(a) (IPTA, SKA-5yrs, combined): $f_0 = (5.6,14,8.1)\times10^{-9}$ Hz, $\Omega_0 = 2.9 \times 10^{-10} \pm (11.0,13.4,8.5) \times 10^{-11}$, $m=-0.01 \pm (0.80,1.14,0.47)$.
$f_0 = 7.9\times10^{-9}$ Hz, $\Omega_0 = 2.9 \times 10^{-10} \pm (13.2,13.3,9.4) \times 10^{-11}$, $m=-0.01 \pm (1.09,1.14,0.48$).\\
(b) (IPTA, SKA-20yrs, combined): $f_0 = (5.6,10,8.8)\times10^{-9}$ Hz, $\Omega_0 = 2.9 \times 10^{-10} \pm (11.0,5.8,5.2) \times 10^{-11}$, $m=-0.01 \pm (0.80,0.29,0.26)$.}
\label{fig:overlapping}
\end{figure}

A useful test is then whether the new experiment renders its predecessor irrelevant with regards to the \ac{SGWB}, or whether there is a noticeable improvement in constraining power when combining both experiments. Figure~\ref{fig:overlapping} considers the combined sensitivity of \ac{IPTA}+\ac{SKA} to a fiducial model representing \acp{PBH} (PBHs2, \cite{PBHs2}) in two cases: where SKA has been observing for 5 years, and for the full 20 year observing run. When the two experiments' individual sensitivities are at similar levels, the inclusion of \ac{IPTA} reduces the \ac{SKA} errors $\sigma_{Y_0}$ and $\sigma_m$ by factors of 1.4 and 2.4, respectively. But by the time the sensitivity of SKA reaches even one decade lower, the improvements added by IPTA are negligible.

Similarly, and now considering a fiducial model of cosmic strings with tension of $G\mu \approx 10^{-11}$ (strings1b, \cite{strings1}), the less sensitive LVK-O4 can only offer \ac{ET} neglible reductions in $\sigma_{Y_0}$ and $\sigma_m$. Whereas \ac{CE} with its similar sensitivity to \ac{ET} confers reductions of 1.5 and 2.1, for cosmic strings with a tension of $G\mu \approx 10^{-15}$ (strings1d, \cite{strings1}).

Per the Fisher matrix formalism, two mathematically identical experiments run in conjunction would give an improvement of $\sqrt{2}\approx 1.4$ over each experiment individually. \ac{CE} and \ac{ET} are not identical, and in fact give a reduction in $\sigma_{Y0}$ of \textit{at least} $\sqrt{2}$ in combination with each other, but give even more improved results for $\sigma_m$. This is largely as a result of the widening of the most sensitive frequency range.

These results are generalisable: a new experiment with better sensitivity does not gain significant improvements to its constraining power with the retention of its predecessor,\footnote{Although it should be noted that there are other science cases for continuing observations with older experiments, including their use in locating (non-\ac{SGWB}) sources on the sky, given sufficiently high \ac{SNR}, and the mitigation of systematic error.} but combinations of experiments with similar sensitivities can be very useful in reducing the errors. Future experiments with similar sensitivities and frequency ranges are planned to run in conjunction: \ac{CE}+\ac{ET} (which are specifically intended to run as a combined experiment, like \ac{LVK}, to obtain better spatial resolution on individual sources), and \ac{BBO}+\ac{DECIGO} (which are not planned as a combined experiment, but do have significant frequency overlap). 

\subsubsection{Distinct Experiments}
With this confirmation that overlapping experiments confer additional sensitivity only when one is not much more sensitive than the other, we now turn to combinations of experiments which do not overlap.

Experiments covering a wide range of frequency space will give information on the overall shape of the \ac{SGWB} spectrum, even individually. We look to see what the improvements will be when experiments at different frequencies come online and are evaluated in combination. In particular, we consider groups of experiments which straddle gaps in observing frequency, to establish whether the improved sensitivities of experiments in combination can alleviate issues around unobserved frequencies.

The results are highly dependent on the fiducial model---a \ac{SGWB} spectrum which is already very well constrained by a single experiment, but lies beneath the sensitivity region of another, is no better constrained by the combination of experiments. The cases where evaluating experiments in combination works well are those where the fiducial model is reasonably constrained by each experiment independently.

\begin{figure}[H]
\centering
\begin{subfigure}[b]{\textwidth}
\centering
\includegraphics[width=0.49\textwidth]{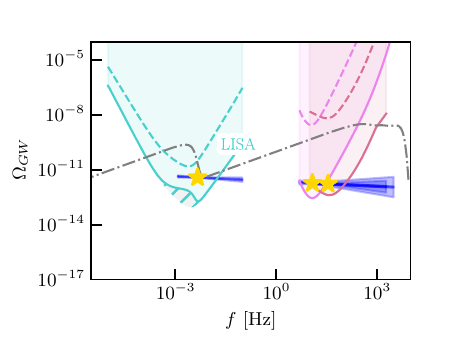}
\includegraphics[width=0.49\textwidth]{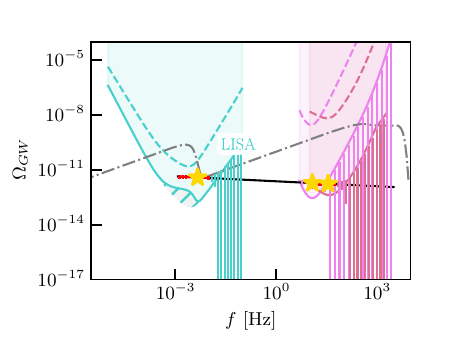}
\caption{}\label{subfig:distinctA}
\end{subfigure}
\hfill
\begin{subfigure}[b]{\textwidth}
\centering
\includegraphics[width=0.49\textwidth]{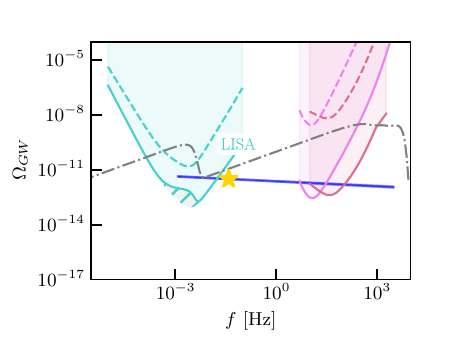}
\includegraphics[width=0.49\textwidth]{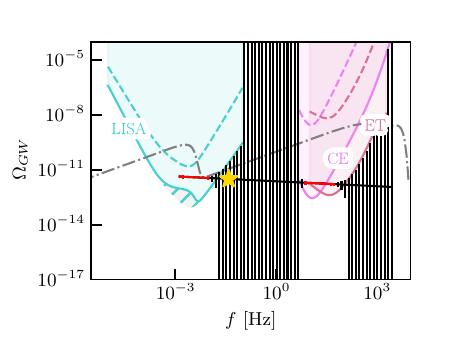}
\caption{}\label{subfig:distinctB}
\end{subfigure}
\caption{The sensitivities of \ac{ET}, \ac{CE} and \ac{LISA} to (strings1d), shown both as experiments operating individually and as a group.\\
(a) (LISA,ET,CE): $f_0 = (0.005,12.1,35.5)$ Hz, $\Omega_0 = (3.9,1.9,1.7) \times 10^{-12} \pm (1.2,1.3,1.4) \times 10^{-13}$, $m=-0.09 \pm (0.08,0.22,0.17)$. Range constrained to 1\%: (0.001--0.010, 8.13--21.4, 19.3--55.1) Hz.\\
(b) LISA+CE+ET combined: $f_0 = 0.04$ Hz, $\Omega_0 = 3.2 \times 10^{-12} \pm 8.4 \times 10^{-14}$, $m=-0.09 \pm 0.007$. Range constrained to 1\%: 0.001--0.1, 8.13--55.1 Hz.}
\label{fig:distinct}
\end{figure}

The example shown in figure \ref{fig:distinct} compares the individual and combined sensitivities of \ac{LISA}, \ac{CE} and \ac{ET} to a fiducial model representing cosmic strings with tension of $G\mu \approx 10^{-15}$ (strings1d, \cite{strings1}). Each experiment individually constrains $\Omega$ by a similar amount, $\sigma_\Omega \approx 1.3\times 10^{-13}$, but \ac{LISA} constrains $m$ considerably better than the ground-based experiments, with $\sigma_{m\textrm{ LISA}} = 0.08$ compared to $\sigma_{m\textrm{ CE, ET}} \approx 0.2$. These constraints reduce to $\sigma_\Omega \approx 8.4\times 10^{-14}$ and $\sigma_{m} = 0.007$, improvements of 1.5 and 11, respectively.

In general, very good constraints can be conferred from one end of the spectrum to the other, but this relies on having confidence in the fiducial model as a single power law. In cases where we are confident in this shape of power law across a wide frequency range (e.g., some models of inflation and cosmic strings, or the $f^{2/3}$ sections of astrophysical foregrounds), this is reasonable. However, while experiments can work together across a frequency gap to constrain a fiducial model, they cannot generate any errors bars in the `unseen' frequency bands between experiments (see the right-hand panels in figure \ref{fig:distinct}).

\subsection{NANOGrav15 as a Prior}
The addition of the NANOGrav15 results as a prior behaves identically to the addition of a distinct experiment, although any variation in the fiducial model compared to the posteriors from NANOGrav15 must be dealt with as laid out in section~\ref{SS:prior}. And as justified in that section, we only consider fiducial models which lie within a reasonable distance from the posteriors. In the case of a single power law, this is particularly limiting, although this requirement is relaxed when more complicated spectral shapes are introduce (and will be considered in our upcoming paper \cite{paper2}.

\begin{figure}[H]
\centering
\begin{subfigure}[b]{0.49\textwidth}
\centering
\includegraphics{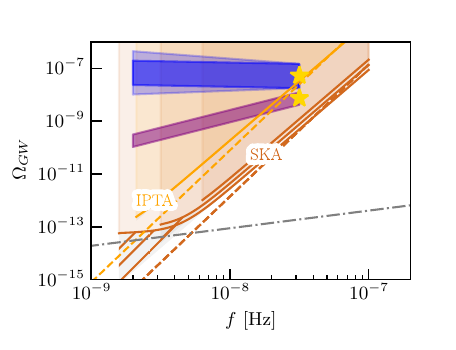}
\caption{}\label{subfig:NG1}
\end{subfigure}
\hfill
\begin{subfigure}[b]{0.49\textwidth}
\centering
\includegraphics{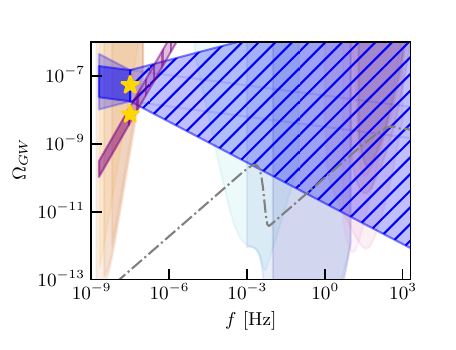}
\caption{}\label{subfig:NG2}
\end{subfigure}
\caption{The NANOGrav15 results \cite{NG15_gwb} plotted over GW experiment sensitivity regions, with the instantaneous foreground, for comparison. The blue shaded region shows the joint $A_\textrm{CP}$-$\gamma_\textrm{CP}$ posterior, and the purple region the $A_\textrm{CP}$ posterior for fixed $\gamma_\textrm{CP}=13/3$, both to the 95\% CL.}
\end{figure}

The NANOGrav15 results \cite{NG15_pulsars,NG15_gwb} are given in both joint $A_\textrm{CP}$-$\gamma_\textrm{CP}$ posteriors, and a single $A_\textrm{CP}$ posterior for fixed $\gamma_\textrm{CP}=13/3$ (to match the expected spectral shape of \ac{SMBHBs}). At the reference frequency $f_\textrm{ref}=1$ yr$^{-1}$ the posteriors are independent, and can be converted to decorrelated errors on $Y_0$ and $m$. The two sets of posteriors (with and without fixed $\gamma_\textrm{CP}$) converted into our $Y_0$ and $m$ parameters with standard linear error propagation are plotted in figure~\ref{subfig:NG1}. The hatched region in figure~\ref{subfig:NG2} then shows an  extrapolation of the posteriors to higher frequencies, assuming an unrealistic spectrum characterized by a single power law at all frequencies.

\begin{figure}
\begin{subfigure}[b]{0.49\textwidth}
\centering
\includegraphics{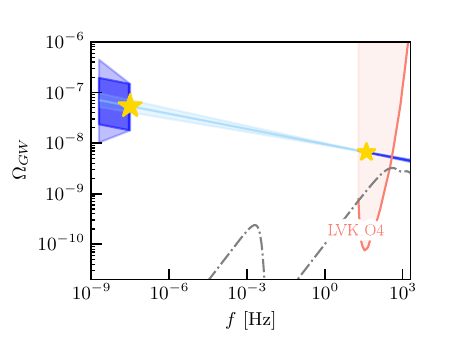}
\caption{}\label{subfig:NG3}
\end{subfigure}
\begin{subfigure}[b]{0.49\textwidth}
\centering
\includegraphics{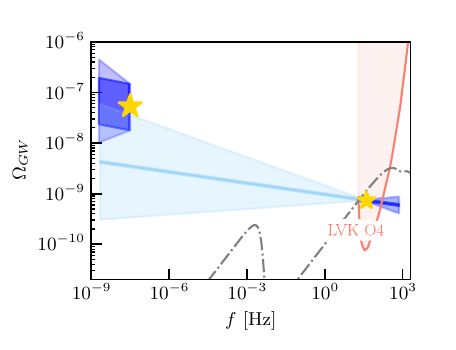}
\caption{}\label{subfig:NG4}
\end{subfigure}
\caption{The sensitivity of LVK-O4, with and without NANOGrav15 as a prior.\\
(a) LVK-O4(+NANOGrav15): $f_0 = 40.4$ Hz, $\Omega_0 = 6.6 \times 10^{-9} \pm 3.12 \ (3.12) \times 10^{-11}$, $m=-0.1 \pm 0.01 \ (0.01)$.\\
(b) LVK-O4(+NANOGrav15): $f_0 = 40.8$ Hz, $\Omega_0 = 7.3 \ (7.1 )\times 10^{-9} \pm 3.12 \ (3.11) \times 10^{-11}$, $m=-0.07 (-0.07) \pm 0.12 \ (0.11)$.}
\label{fig:NG15}
\end{figure}

Figure~\ref{subfig:NG3} shows the addition of the NANOGrav15 results as a prior for LVK-O4, for a fiducial model which matches the NANOGrav15 posteriors exactly. This represents a scenario where the \ac{SGWB} spectrum presents as a single power law from \ac{PTA} to ground-based frequencies, and where it is also detected by the current operational phase of \ac{LVK}. In this case, the inclusion of NANOGrav15 as a prior does not confer any improvements to the constraints that LVK-O4 would set, and indeed LVK-O4 would improve upon the NANOGrav15 constraints themselves.

A different case, if LVK-O4 were to make observations of a \ac{SGWB} spectrum which deviated from the NANOGrav posteriors, is shown in figure~\ref{subfig:NG4}. In this scenario, per the discussion in section \ref{SS:prior}, the combination of the two initial power laws (the NANOGrav15 prior, and that observed by LVK-O4) can be thought of as constraints on a third power law spectrum, consistent with both sets of observations. Again in this case, the inclusion of NANOGrav15 as a prior only brings negligble improvements to the constraints.

\section{Discussion}\label{Se:Discussion}
Following on from the first detections of the \ac{SGWB} at \ac{PTA} frequencies, and in advance of any detections at other frequencies, we remain reliant on upper limits and forecast sensitivities in our understanding of the wider \ac{SGWB} spectrum.

Previous works have primarily focused on the sensitivities of given experiments to one, or a range of, cosmological \ac{SGWB} spectra (see, e.g., \cite{Gowling_2021,Boileau_2023,Gowling_2023,cecchini2025forecastingconstraintssigwfuture,wang2025detectingstochasticgravitationalwave,caporali2025impactcorrelatednoisereconstruction,babak2024forecastingsensitivitypulsartiming,pol2023lisagammaraytelescopesmultimessenger,Caprini_2019,tan2024}). A handful of works have considered the improvement to LISA's sensitivity when a second experiment is combined \cite{torresorjuela2024,marriottbest2024exploringcosmologicalgravitationalwave,wang2022probinggravitationalwavebackground}, and ref. \cite{Braglia_2021} discusses the  benefit of multi-frequency detections to an example PBH spectrum. Whereas in this paper we generalise to evaluating the respective sensitivities of current and upcoming \ac{GW} experiments to \ac{SGWB} spectra which map to generic single power laws, looking specifically at the improvement to sensitivity which is conferred by utilising multiple experiments in conjunction with each other in the presence of realistic foreground contamination. 
Our results are given in terms of the constraints that the experimental setup confers on the amplitude and slope of the fiducial model, and the subsequent improvements on these when experiments are evaluated in combination with each other.  We plot these as `bowtie' shapes, which are a reasonable approximation to the true combination of amplitude and slope errors.

These bowtie shapes are a useful visualisation of the results, for the region where they lie above our $\Omega_\textrm{nwt}$ curves. As discussed in section~\ref{Results-indiv}, they can over- and underestimate the error bars below the curves.

Our generalisable findings, discussed further in section~\ref{Results-overl}, reveal that a more sensitive replacement experiment does not gain noticeable improvement in its constraints on the slope or amplitude from the retention of its predecessor, while the inclusion of another experiment at similar sensitivity levels often shows greater than the simplest $\sqrt{2}$ improvement.

We also show that the combination of experiments across a gap in observing frequency can confer good improvements to their individual constraints, under the assumption of a continuous fiducial model across the frequencies in question. We demonstrate the inclusion of the most recent NANOGrav results as a prior for fiducial models, and show that they do not bring a noticeable improvement to signals detectable by LVK-O4.
    
A future paper \cite{paper2} is in preparation, extending this work to include single and double broken power laws, which provide a representative fit of a wider range of predicted \ac{SGWB} spectra. By extending the range of predicted \ac{SGWB} spectra which can act as fiducial models, we will then be able to use the techniques in this work to construct a timeline for when it might be possible to make observations of different contributions of the \ac{SGWB}.
 
\appendix

\section{Sensitivity Curves for GW Experiments}\label{A:noise}

Given below are the nominal sensitivities $\Omega_\textrm{nom}$ of the experiments considered in this work. These are given in terms of $S_\textrm{nom}(f)$, and require conversion to $\Omega_\textrm{nom}$ through eq.~\ref{eq:ins_noise}. Where simple analytic approximations to the instrumental noise have been used, they are given below. Otherwise, references with more complicated fits are listed.

\subsection{Ground-Based Interferometers}
\begin{itemize}
    \item \textbf{Ligo-Virgo-Kagra:}

Strain curve information can be found in reference \cite{LVK_curves}. 
    \item \textbf{Cosmic Explorer:} Strain curve information can be found in reference \cite{CE_curves}. 
    \item \textbf{Einstein Telescope:} Strain curve information can be found in reference \cite{CE_curves}. 
\end{itemize}

\subsection{Space-Based Interferometers}
\begin{itemize}
    \item \textbf{LISA:} The strain is given by
\begin{equation}
    S = \frac{40}{3} \times \left[\frac{S_1(f)}{(2\pi f)^4} + S_2(f) \right]\times \left( 1+\left(\frac{3f}{4f_t}\right)^2\right), 
\end{equation}
where $f_t = (3\times10^8\, \textrm{ms}^{-1})/(2\pi(2.5\times10^8\, \textrm{m}))\approx 0.2 \, \textrm{Hz}$, $S_1 = 5.76\times10^{-48} \, \textrm{s}^{-4} \, \textrm{Hz}^{-1}\times(1+(0.4\, \textrm{mHz}/f)^2) \, \textrm{Hz}^{-1}$ is the single-link optical metrology noise, and $S_2 =3.6\times10^{-41}\,\textrm{Hz}^{-1}$ is the single test mass acceleration noise \cite{Robson_2019,Gowling_2021,LISA_doc}.
    \item \textbf{DECIGO:} The strain is given by
\begin{equation}
\begin{split}
S &= \left[7.05\times10^{-48}\times\left(1+\left(\frac{f}{7.36\; \textrm{Hz}}\right)^2\right)\right]
+\left[\frac{4.8\times10^{-51}}{1+\left(\frac{f}{7.36\; \textrm{Hz}}\right)^2} \times \left(\frac{f}{1\; \textrm{Hz}}\right)^{-4}\right]\\
&\phantom{=}+\left[5.33\times10^{-52}\times \left(\frac{f}{1\; \textrm{Hz}}\right)^{-4}\right]\; \textrm{Hz}^{-1},
\end{split}
\end{equation}

where the first term shows the shot noise, the second the radiation pressure noise, and the third the acceleration noise \cite{Yagi_2011}.
    \item \textbf{BBO:} The strain is given by 
\begin{equation}
S = \left[2.00\times10^{-49}\left(\frac{f}{1^{-4}}\right)^2 + 4.58\times10^{-49}+1.26\times10^{-51}\left(\frac{f}{1^{-4}}\right)^{-4}\right]\;\textrm{Hz}^{-1},  
\end{equation}
 where the three terms likewise represent the shot, radiation pressure, and acceleration noises \cite{Yagi_2011}.
\end{itemize}

\subsection{Pulsar Timing Arrays}
The strain of a pulsar timing array is given by
\begin{equation}\label{eq:PTA_noise}
    S = \sqrt{\frac{2}{N \left(N-1\right)}}\times\frac{D}{\zeta_\textrm{rms} R},
\end{equation}
where $N$ is the number of pulsars observed, $D = 2t\sigma^2$ is the intrinsic (white) timing noise per pulsar, $t$ is the average cadence of observations, $\sigma$ is the standard deviation of the detector noise, $\zeta_\textrm{rms}=0.147$ is the rms value for the Hellings-Downs factor \cite{Hellings_Downs} between two pulsars, and $R=1/(12\pi^2f^2)$ is the detector response function \cite{curves}. 
\begin{itemize}
    \item \textbf{IPTA:} The strain is given by equation~\ref{eq:PTA_noise},
with $N=20$, $t=12.08\times10^5\,$s, $\sigma=10.0\times10^{-8}\,$s \cite{curves}.
    \item \textbf{SKA:} The strain is given by equation~\ref{eq:PTA_noise},
with $N=50$, $t=6.04\times10^5\,$s, $\sigma=3.0\times10^{-8} \,$s \cite{curves}.
\end{itemize}

\section{Fiducial Models}\label{A:fiducial_models}

\begin{center}
\begin{longtable}{lm{4.5cm}rrcrc}
\caption{}
\label{tab:fiducials} \\

\hline \multicolumn{1}{c|}{Model} & \multicolumn{1}{c|}{Details} & \multicolumn{1}{c|}{$f_1$} & \multicolumn{1}{c|}{$f_2$} & \multicolumn{1}{c|}{$\Omega_2$} & \multicolumn{1}{c|}{$m$} & \multicolumn{1}{c}{Ref.} \\ \hline 
\endfirsthead

\endhead

\hline \multicolumn{7}{r}{{Continued on next page}} \\ \hline
\endfoot

\endlastfoot

\multicolumn{7}{c}{Inflation:} \\
\hline
        inflation1 & \ac{SFSR}, $r=0.1$ & $-9.94$ & $2.42$ & $1.76\times 10^{-16}$ & $0.00$ & \cite{inflation1}\\
        inflation2 & \ac{SFSR}, $r=0.2$ & $-9.94$ & $3.07$ & $2.64\times 10^{-16}$ & -0.02 & \cite{inflation2}\\
\hline
\multicolumn{7}{c}{Primoridal Black Holes:} \\
\hline
        PBHs1 & solar mass \acp{PBH}
        & \parbox{.8cm}{\centering --} & \parbox{.8cm}{\centering --} & -- & \parbox{.8cm}{\centering --} & \cite{PBHs1}\\
        PBHs2 & \acp{PBH} which would satisfy all of dark matter
        & $-9.02$ & $-1.25$ & $2.47\times 10^{-10}$ & -0.01 & \cite{PBHs2}\\
\hline
\multicolumn{7}{c}{Cosmic Strings:} \\
\hline
        strings1a & $G\mu \approx 10^{-9}$ & $-8.84$ & $0.78$ & $6.45\times 10^{-10}$ & -0.07 & \cite{strings1}\\
        strings1b & $G\mu \approx 10^{-11}$ & $-7.11$ & $4.30$ & $6.17\times 10^{-11}$ & -0.07 & \cite{strings1}\\
        strings1c & $G\mu \approx 10^{-13}$ & $-4.76$ & $4.15$ & $7.55\times 10^{-12}$ & -0.08 & \cite{strings1}\\
        strings1d & $G\mu \approx 10^{-15}$ & $-2.92$ & $4.50$ & $9.24\times 10^{-13}$ & -0.09 & \cite{strings1}\\
        strings1e & $G\mu \approx 10^{-17}$ & $-0.87$ & 3.82 & $9.63\times 10^{-1}$ & -0.12 & \cite{strings1}\\
        strings1f & $G\mu \approx 10^{-19}$ & $-0.87$ & 3.82 & $9.63\times 10^{-1}$ & -0.12 & \cite{strings1}\\
        strings1g & $G\mu \approx 10^{-21}$ & $-0.87$ & 3.82 & $9.63\times 10^{-1}$ & -0.12 & \cite{strings1}\\
        strings1h & $G\mu \approx 10^{-23}$ & $-0.87$ & 3.82 & $9.63\times 10^{-1}$ & -0.12 & \cite{strings1}\\
        strings2 & secondary pop. of small loops & $0.53$ & $2.13$ & $9.68\times 10^{-8}$ & 0.43 & \cite{strings4}\\
        strings3i & $G\mu \approx 10^{-7}$ (segment 1) & $-5.78$ & $-0.56$ & $7.53\times 10^{-9}$ & -0.38 & \cite{Siemens_2007}\\
        strings3ii & $G\mu \approx 10^{-7}$ (segment 2) & $0.00$ & $3.76$ & $5.02\times 10^{-9}$ & -0.01 & \cite{Siemens_2007}\\
\hline
\multicolumn{7}{c}{First-Order Phase Transitions:} \\
\hline
        FOPTs1a & electroweak ($\approx$200 GeV) & $-9.99$ & $-7.13$ & $4.68\times 10^{-7}$ & 2.83 & \cite{FOPTs1}\\
        FOPTs1b & QCD ($\approx$200 MeV) & $-3.64$ & $-1.19$ & $1.00\times 10^{-16}$ & -3.85 & \cite{FOPTs1}\\
        FOPTs2a & QCD, $\beta/H = 10^2$& $-2.07$ & $0.41$ &$1.40\times 10^{-18}$ & -1.53 & \cite{FOPTs2}\\
        FOPTs2b & QCD, $\beta/H = 10^3$& $-1.52$ & $0.20$ &$1.01\times 10^{-18}$& -1.38& \cite{FOPTs2}\\
        FOPTs2c & QCD, $\beta/H = 10^4$& $-1.44$ &  $-0.61$ & $2.90\times 10^{-18}$& -0.47& \cite{FOPTs2}\\
        FOPTs3ai & $T_n = 10^7$ GeV (segment 1) & $-0.09$ &  $0.76$ & $2.31\times10^{-9}$& $2.85$& \cite{FOPTs3}\\
        FOPTs3aii & $T_n = 10^7$ GeV (segment 2)& $1.30$ &  $3.63$ & $1.30\times10^{-11}$& $-0.99$& \cite{FOPTs3}\\
        FOPTs3bi &$T_n = 10^8$ GeV (segment 1) & $0.71$ &  $1.68$ & $1.99\times10^{-9}$& $2.80$& \cite{FOPTs3}\\
        FOPTs3bii &$T_n = 10^8$ GeV (segment 2) & $2.29$ &  $4.70$ & $1.21\times10^{-11}$& $-1.00$& \cite{FOPTs3}\\
        FOPTs3ci & $T_n = 10^3$ GeV (segment 1)& $-3.60$ &  $-3.08$ & $1.37\times10^{-9}$& $2.63$& \cite{FOPTs3}\\
        FOPTs3cii &$T_n = 10^3$ GeV (segment 2) & $-2.60$ &  $-0.47$ & $1.21\times10^{-11}$& $-1.01$& \cite{FOPTs3}\\
\hline
\multicolumn{7}{c}{Exotic Phenomena:} \\
\hline
        axions1i & Axion-based inflation (segment 1) & $-3.68$ & $-1.33$ & $2.04\times 10^{-11}$ & 1.18 & \cite{axions1}\\
        axions1ii & Axion-based inflation (segment 2) & $-0.12$ & $2.49$ & $1.22\times 10^{-8}$ & 0.47 & \cite{axions1}\\
        \hline
\caption{Details of the fiducial models presented in figure \ref{Fig:example_spectra}, and used as fiducial models in the results. The values listed for $f_1= \log_{10}(f_\textrm{min} / \textrm{Hz})$, $f_2= \log_{10}(f_\textrm{max} / \textrm{Hz})$ correspond to the frequency range over which these models can be usefully approximated by a single power law, where applicable. $\Omega_2 = \Omega(f_\textrm{max})$ is then the power at $f_\textrm{max}$, and $m$ the power law index. All values are approximate.}
\end{longtable}
\end{center}

\section{Foreground}\label{A:foreground}
We model the astrophysical foregrounds as 
\begin{equation}\label{eq:fg_App}
    \Omega_\textrm{fg} = A f ^{2/3} \left[ \ \exp \left( - \left(f/f_\textrm{ref}\right)^B  \right) + D \ \sin^2 \left( \pi f / f_\textrm{ref} \right)\right] \times \left[ 1 + \tanh \left( E \left( f_* - f\right) \right) \right],
\end{equation}
with the coefficients for each binary population given in table~\ref{tab:fgs}. Eq.~\ref{eq:fg_App} is an adaptation of a WD-WD model developed by Cornish and Robson \cite{Cornish_2017}, such that it gives a more general fit across the different binary populations.

\begin{table}[H]
    \centering
    \begin{tabular}{c|ccccccc}
        \hline
        Pop. & $A$ [Hz$^{-2/3}$]&$B$ &$D$ & $E$ [Hz$^{-1}$]& $f_*$ [Hz] & $f_\textrm{ref}$ [Hz] & Model\\
        \hline
        WD-WD & $9.48\times10^{-9}$ & 2.25 & 0.00449 & 8527 & 0.00346 & 0.00449 & \cite{Cornish_2017,Carter_2012}\\
        BH-BH &$6.85\times10^{-11}$ & 2.69& 0.00148 & 0.00139 & 42.5 & 793 & \cite{Capurri_2022,LVK_2021,LVK_2021b}\\
        BH-NS & $7.34\times10^{-12}$ &4.08 & 2.96e-08 &  0.00139 & 1571 & 2771& \cite{Capurri_2022,LVK_2021,LVK_2021b}\\
        NS-NS &$6.89\times10^{-12}$& 3.83 & 0.00242 & 0.000648 & 6481 & 5961 & \cite{Capurri_2022,LVK_2021,LVK_2021b}\\
        \hline
    \end{tabular}
    \caption{Coefficients for the foreground model used in this work, for white dwarf (WD), black hole (BH), neutron star (NS), and BH-NS binaries. These were found with a least-squares fitting to the population model spectra given in the right-hand column: WD-WD from the Sloane Digital Sky Survey \cite{Carter_2012}, and BH-BH, BH-NS, NS-NS from LVK data \cite{LVK_2021,LVK_2021b}.}
    \label{tab:fgs}
\end{table}

\acknowledgments
This work was supported by the Science and Technology Facilities Council.

\bibliographystyle{JHEP}
\bibliography{biblio.bib}

\end{document}